\documentclass[11pt]{article}

\makeatletter
\@addtoreset{equation}{section}
\makeatother

\usepackage{epsfig,amsfonts}
\usepackage[fleqn]{amsmath}
\usepackage{amsthm,amssymb}
\usepackage{yfonts,mathrsfs}
\usepackage{graphicx}
\usepackage{hhline}
\usepackage{cite}
\usepackage{upgreek}
\def\be{\begin{equation}}
\def\ee{\end{equation}}
\def\bal{\begin{align}}
\def\eal{\end{align}}
\def\bea{\begin{eqnarray}}
\def\eea{\end{eqnarray}}
\def\be{\begin{equation}}
\def\ee{\end{equation}}
\def\bdm{\begin{displaymath}}
\def\edm{\end{displaymath}}
\def\bea{\begin{eqnarray}}
\def\eea{\end{eqnarray}}

\def\XXint#1#2#3{{\setbox0=\hbox{$#1{#2#3}{\int}$}
    \vcenter{\hbox{$#2#3$}}\kern-.5\wd0}}

\begin{document}

\begin{titlepage}
\begin{flushright}
RUNHETC-2013-20\\
\end{flushright}

\vspace{1.2cm}

\begin{center}
\begin{LARGE}
{\bf Ising Spectroscopy II:\\ Particles and poles at $T > T_c$.}

\vspace{0.2cm}

\end{LARGE}

\vspace{1.2cm}

\begin{large}

{\bf A.B. Zamolodchikov}$^{1,2}$

\end{large}

\vspace{1.cm}

{${}^{1}$NHETC, Department of Physics and Astronomy\\
     Rutgers University\\
     Piscataway, NJ 08855-0849, USA\\

\vspace{.2cm}

${}^{2}$Institute for Information Transmission Problems\\
Moscow 127994, Russia\\
 }

\vspace{1.4cm}

\centerline{\bf Abstract} \vspace{.8cm}
\parbox{11cm}{
I discuss particle content of the Ising field theory (the scaling limit of
the Ising model in a magnetic field), in particular the evolution of its mass
spectrum under the change of the scaling parameter. I consider both real and
pure imaginary magnetic field. Here I address the high-temperature regime,
where the spectrum of stable particles is relatively simple (there are from
one to three particles, depending on the parameter). My goal is to understand
analytic continuations of the masses to the domain of the parameter where
they no longer exist as the stable particles. I use the natural tool -- the
$2\to 2$ elastic scattering amplitude, with its poles associated with the
stable particles, virtual and resonance states in a standard manner.
Concentrating attention on the "real" poles (those corresponding to stable
and virtual states) I propose a scenario on how the pattern of the poles
evolves from the integrable point $T=T_c,\ H\neq 0$ to the free particle
point $T>T_c,\ H=0$, and then, along the pure imaginary $H$, to the Yang-Lee
critical point. Waypoints along this evolution path are located using TFFSA
data. I also speculate about likely behavior of some of the resonance poles.}
\end{center}

\bigskip

\begin{flushleft}
\rule{4.1 in}{.007 in}\\
{October  2013}
\end{flushleft}
\vfill

\end{titlepage}
\newpage

\section{Introduction}

This paper is the second part of the project "Ising Spectroscopy" devoted to
detailed study of the particle mass spectrum in the Ising field theory. The
latter is the quantum field theory of the scaling domain of 2D Ising model in
a magnetic field. It can be defined via the formal action
\begin{eqnarray}\label{ift}
{\cal A}_{\rm IFT} = {\cal A}_{\rm c={1/2}\ CFT}\, +\,
\frac{m}{2\pi}\int\varepsilon (x)\,d^2 x\, + \,h \int\sigma(x)\,d^2 x\ .
\end{eqnarray}
Here $\cal{A}_{\text{c=1/2\ CFT}}$ represents the unitary conformal field
theory with the central charge $1/2$ (which of course is the theory of free
massless Majorana fermions), while $\varepsilon(x)$ and $\sigma(x)$ are two
relevant primary fields of this CFT, with the conformal dimensions
$(1/2,1/2)$ and $(1/16,1/16)$, respectively (see e.g. \cite{itzykson,cft}).
In terms of the Ising model, the parameter $h$ in \eqref{ift} is suitably
scaled magnetic field, while $m \sim T_c - T$ describes the temperature
deviation from the Curie point. Away from the critical point $m =0,\ h = 0$
the theory \eqref{ift} is massive, and therefore much of its physical content
can be understood through its spectrum of particles, stable ones and
resonances. Up to overall scale, physical content of the theory \eqref{ift}
is controlled by a single dimensionless parameter
\begin{figure}[ht]
\centering
\includegraphics[width=13cm]{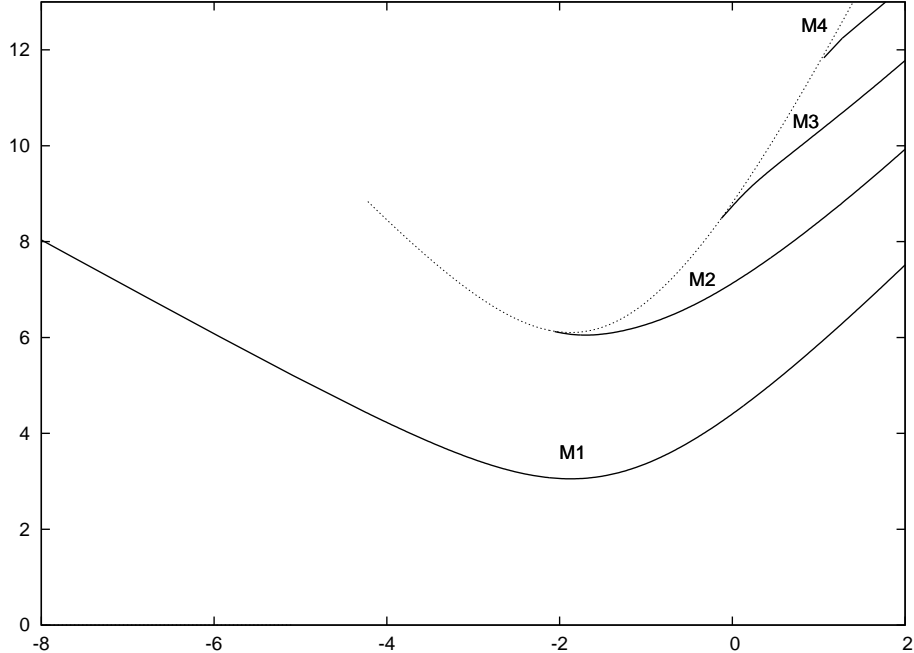}
\caption{\small{The lowest masses $M_n$ as the functions of $\eta$. The dotted
line shows the stability threshold $2 M_1$. As $\eta$ decreases the
particles successively disappear from the spectrum, becoming virtual, and then
resonance states (see Sect 3).}}\label{masses}
\end{figure}
\begin{eqnarray}\label{etadef}
\eta = \frac{2\pi\,m}{\ \,|h|^{8/15}}\,.
\end{eqnarray}
In particular, the number of stable particles and their masses $M_n$ change
with this parameter. Figure 1 shows the behavior of few lowest masses
$M_n(\eta)$ (measured in the units of $|h|^\frac{8}{15}$) at real $\eta$.
This picture (obtained numerically in \cite{fz1}) substantiates the scenario
originally proposed in Ref. \cite{McCoy}: when $\eta$ changes from $-\infty$
to $+\infty$ the particle spectrum evolves from a single particle to an
infinite tower of "mesons" formed by weakly confined "quarks". In the
process, the mass spectrum exhibits rather intricate behavior. Understanding
the mass spectrum at intermediate (and not less importantly, complex) values
of $\eta$ is the main goal of this project.

In the first part, Ref.\cite{fz3}, the focus was set on the "low-T" domain
$\eta >0$, and an attempt was made to understand details of the mass spectrum
there in terms of confined "quarks". Although many interesting questions
about the particle spectrum in that domain (especially the questions
regarding how the masses behave near the stability thresholds, and their fate
as the resonance states) were left open, here instead we make a foray into
the "high-T" domain $\eta \leq 0$. In this region the spectrum of stable
particles is actually relatively simple - there is always the lightest stable
particle which we denote $A_1$ (and its mass $M_1$), and at sufficiently
large $\eta$ in addition the heavier particles $A_2$ and $A_3$ appear. The
behavior of the masses $M_1$, $M_2$ and $M_3$ as the functions of $\eta$ is
shown in Fig.1. However, it is interesting to understand the analytic
continuations of the functions $M_n(\eta)$ to the region where they no longer
represent the masses of stable particles.

Convenient tool for addressing this question is the elastic $A_1 A_1 \to A_1
A_1$ scattering amplitude $S(\theta)$. It is a complex analytic function of
the rapidity difference $\theta$ of the colliding particles, and its poles in
this variable are associated with either stable particles of the theory or
virtual or resonance states. In this work I address the question how the
pattern of the poles of $S(\theta)$ evolves as $\eta$ changes form $0$ to
$-\infty$. In fact, the domain of negative real $\eta$ represents only a part
of interesting high-T region of parameters where the theory \eqref{ift} is
"real": at pure imaginary $h$ below the so called Yang-Lee critical point
\cite{Fisher} the vacuum energy density and the particle mass $M_1$ remain
real, and despite the fact that the theory is no longer unitary, it is still
meaningful (and interesting) to study its S-matrix. Therefore, I will
actually discuss the evolution of the pattern of poles in the wider region,
when the parameter $h^2/(-m)^\frac{15}{4}$ changes from $+\infty$ to $0$, and
then decreases further from $0$ to the Yang-Lee point $-\xi_{0}^2=-
0.035846...\ $. In this work I will concentrate attention on the "real" poles
of $S(\theta)$, the ones that correspond to the stable particles and virtual
states, because the real poles usually dominate the low-energy behavior of
the scattering, and also because they are much easier to analyze.
Nonetheless, remarks about some of the resonance poles will be made in
Sect.4.

My analysis will be based on general principles, such as positivity of
residues of the particle poles in the unitary domain of \eqref{ift}, as well
as a combination of analytical and numerical data about the mass spectrum of
\eqref{ift}. The analytical data include exact mass spectra at the integrable
points \cite{itzykson,e8,cardy-mussardo}, and the results of the perturbation
theory expansions around them \cite{Mussardo1,e8decay,delfino2,ZZy}.
Numerical data is obtained by using the Truncated Free Fermion Space Approach
(TFFSA) of \cite{fz1} (which is an adaptation of the TCSA of \cite{alz1} to
the theory \eqref{ift}). In this work I only use the numerical data on the
masses of the stable particles, which is extracted from the finite-size
energy spectra numerically evaluated via the TFFSA \cite{fz1}. The numerics
is used as a general guidance, and also for estimating numerical values of
the parameter \eqref{etadef} associated with important waypoints in
\eqref{ift}. Potentially, TFFSA can be also applied to for numerical
evaluations of the scattering phases \cite{alz2}, and masses and width of
resonances \cite{tcsares}. This important task goes beyond the scope of this
work.

\section{$2\to 2$ S-matrix element. Generalities}

Since \eqref{ift} has the particle $A_1$ in its spectrum at all $\eta$, we
find it useful to discuss the other particles in terms of poles in the
S-matrix element $S(\theta)$ associated with the elastic $A_1 A_1 \to A_1
A_1$ scattering. Let us start with brief summary of general analytic
properties of this amplitude\footnote{The content of this section is mostly
an adaptation of the textbook basics of the S-matrix theory (see e.g.
\cite{book}) to 1+1 dimensional kinematics. I include it in order to
introduce suitable notations.} \footnote{Although we explicitly speak about
the Ising field theory \eqref{ift}, the discussion of this section applies to
any scattering theory involving a singlet neutral particle $A_1$.}.

\subsection{Analyticity and Poles}

As usual, it is convenient to characterize the kinematic states of the
particles $A_1$ by their rapidities $\theta$ which parameterize the
two-momenta as $p^\mu = (M_1\,\cosh\theta, M_1 \sinh\theta)$. We will use the
notation $A_1(\theta)$ for the particle with the rapidity $\theta$. The $2\to
2$ S-matrix element $S(\theta)$ is defined as
\begin{eqnarray}\label{s11}
\mid A_1(\theta_1)A_1(\theta_2)\,\rangle_\text{in} = S(\theta_1-\theta_2)
\mid A_1(\theta_1)A_1(\theta_2)\,\rangle_\text{out} + \text{inelastic
terms}\,,
\end{eqnarray}
where the "inelastic terms" include all kinematically admissible states of
$n\geq 3$ particles $A_1$, as well as the states with the bound-state
particles $A_2$, $A_3$, when present in the theory. By standard analyticity,
$S(\theta)$ is analytic in the complex $\theta$-plane, except for poles
(which will be of our primary interest here), and branching points associated
with the inelastic thresholds. The $A_1 A_1 \to X$ thresholds are located at
the points $\pm\,\theta_X + i\pi \mathbb{Z}\,$, where $\theta_X$ are real
positive solutions of the threshold energy equations $2M_1\,\cosh(\theta_X/2)
= E_\text{min}(X)\,,$ $E_\text{min}(X)$ being the minimal energies of
possible combinations $X \neq A_1 A_1$ of stable particles present in the
theory. By introducing the branch cuts from $+\theta_\text{min} + i\pi\,N$ to
$+\infty+i\pi\,N$, and from $-\theta_\text{min}+i\pi\,N$ to
$-\infty+i\pi\,N$, $N \in \mathbb{Z}$, where $\theta_\text{min} = \inf_{X}
\theta_X$, one defines the "principal sheet" of the Riemann surface
associated with $S(\theta)$ \footnote{Certainly, analytic continuation under
the inelastic branch cuts, to further sheets of the Riemann surface is
possible. At the moment I do not have much to say about analytic properties
of the amplitude there.}. Henceforth, speaking of complex $\theta$ we always
refer to the values at the principal sheet. There, the function $S(\theta)$
satisfies the analytic conditions
\begin{eqnarray}\label{crossing}
S(\theta)=S(i\pi-\theta)
\end{eqnarray}
and
\begin{eqnarray}\label{unitarity}
S(\theta)S(-\theta)=1\,.
\end{eqnarray}
The first of them is just the standard expression of the crossing symmetry,
while the second follows from elastic unitarity at real $\theta$ with
$|\theta| < \theta_\text{min}$. Eq's \eqref{crossing} and \eqref{unitarity}
imply periodicity $S(\theta+2\pi\,i)=S(\theta)$, and in what follows we
concentrate attention at the strip
\begin{eqnarray}\label{fullstrip}
-\pi<\Im m\,\theta\leq\pi\,.
\end{eqnarray}
In view of \eqref{unitarity} one may limit attention to the "physical strip"
(PS) $\ 0 \leq \Im m\,\theta \leq \pi$ (which corresponds to the principal
sheet of the Riemann surface for the invariant energy square $s =
4M_{1}^2\,\cosh^2(\theta/2)$), but we find it useful to keep view of the full
strip \eqref{fullstrip}. Since the values of $S(\theta)$ in the strip $-\pi
\leq \Im m\,\theta \leq 0$ are determined through its values in the physical
strip via \eqref{unitarity}, we will refer to it as the "mirror strip" (MS).

In addition to the above analytic conditions, $S(\theta)$ satisfies
\begin{eqnarray}
S(\theta) = S^* (-\theta^*)\,,
\end{eqnarray}
i.e $S(\theta)$ is real analytic function of the variable $\alpha =
-i\theta$. It is often useful to distinguish between the positive and
negative parts of the physical the mirror strips. These are defined as
follows,
\begin{eqnarray}\label{psp}
PS^{(\pm)}: \qquad \ \ \,0  \leq \Im m \,\theta \leq \pi \,, \quad \pm\,\,\Re e\,\theta > 0
\label{psm}
\end{eqnarray}
and
\begin{eqnarray}\label{msp}
MS^{(\pm)}: \qquad -\pi  \leq \Im m \,\theta \leq 0 \,, \quad \pm\,\,\Re e\,\theta > 0\,.
\label{msm}\end{eqnarray}

\begin{figure}[ht]
\centering
\includegraphics[width=12cm]{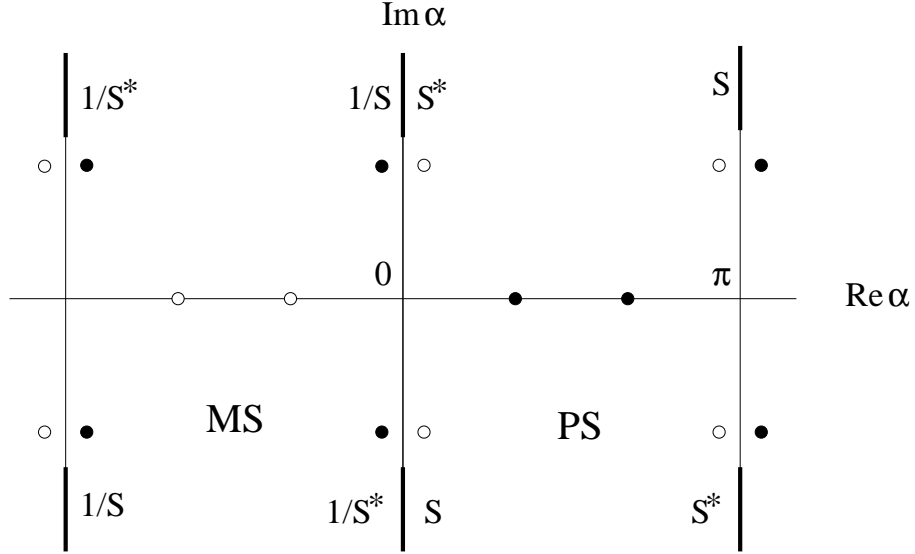}
\caption{\small{Typical analytic structure of the two-particle scattering
amplitude $S(i\alpha)$ in the complex $\alpha$-plane. The bold
lines show the branch cuts associated with inelastic channels.
The values of $S(i\alpha)$ at different edges of the branch cuts
represent physical S-matrix element $S$, its complex conjugate $S^*$,
and the inverse values. The bullets $\bullet$ and circles $\circ$
indicate possible positions of poles and zeroes, respectively. Poles
located on the real $\alpha$-axis, within the physical strip
$0 < \Re e \alpha < \pi$ are associated with the stable particles;
complex poles on the mirror strip  $-\pi<\Re e \alpha<0$ are interpreted
as the resonance scattering states.}
}\label{thetaplane}
\end{figure}

The amplitude $S(\theta)$ may have poles on the principal sheet. According to
\eqref{crossing} and \eqref{unitarity}, it can be generally written as
\begin{eqnarray}\label{poleform}
S(\theta) = \prod_p\,\frac{\sinh\theta + i \sin\alpha_p}
{\sinh\theta - i \sin\alpha_p}\,
\ \exp\big({i\Delta(\theta)}\big)\,,
\end{eqnarray}
where the product factor accounts for all the poles, so that the "inelastic
phase" $\Delta(\theta)$ is analytic everywhere within the principal sheet of
the $\theta$-surface. It takes the real values in the lacunae
$[-\theta_\text{min}: +\theta_\text{min}]$ of the real $\theta$-axis, and and
hence it is a real analytic function of $\theta$ on the principal sheet. The
discontinuity across the inelastic branch cut from $\theta_\text{min}$ to
$\infty$ relates to the "total inelastic cross section"
$\sigma_\text{tot}(\theta)$ \ - \ the total probability of all inelastic
processes in the $A_1 A_1$ scattering at the center of mass energy
$E=2M_1\,\cosh\frac{\theta}{2}$ \ - \  as follows
\begin{eqnarray}
\Delta(\theta+i0)-\Delta(\theta-i0) =
-i\,\,{{\log\big(1-\sigma_{\rm tot}(\theta)\big)}\over{\sinh\theta}}
\end{eqnarray}
This of course is the 1+1 dimensional version of the optical theorem, which
allows one to write down the dispersion relation expressing $\Delta(\theta)$
in terms of $\sigma_\text{tot}(\theta)$. Here I will not discuss the
inelastic factor any further. I only note that in integrable theories all
inelastic processes are forbidden, the function $\Delta(\theta)$ vanishes
identically, and the function $S(\theta)$ reduces to the product factor in
Eq.\eqref{poleform}.

Each factor in the product in \eqref{poleform} is responsible for two poles
on the principal sheet, located at $\theta = i\alpha_p$ and
$i(\pi-\alpha_p)$, $\text{mod} \ 2i\pi \mathbb{Z}$. We denote $r_p$ the
residue of $S(i\alpha)$ at the pole at $\alpha = \alpha_p$,
\begin{eqnarray}
S(\theta) \simeq \frac{ir_p}{\theta - i\alpha_p}\,.
\end{eqnarray}

The parameters $\alpha_p$ are subjects to certain general restrictions. Thus,
the real analyticity of $S(i\alpha)$ as the function of $\alpha$ demands that
$\alpha_p$ are either real or come in complex conjugated pairs; we will refer
to the former as the "real poles" (as opposite to the "complex poles" with
$\Im m\,\alpha_p \neq 0$). The real poles will be of the central interest in
our discussion below. The residues $r_p$ at the real poles are real, and we
will call the the real pole "positive" if $r_p > 0$, or "negative" if $r_p <
0$. As follows from \eqref{crossing}, if the pole at $i\alpha_p$ is positive,
the associated pole at $i(\pi-\alpha_p)$ is negative, and vice versa. For the
complex poles we call the pole at $i\alpha_p^*$ conjugated to the pole at
$i\alpha_p$.

Real poles located in the PS are generally interpreted in terms of the stable
particles of the theory - the $A_1 A_1$ bound states. In unitary field
theories the positivity requires that any positive pole with $\alpha_p \in
[0,\pi]$ is identified with the s-channel bound state with the mass
\begin{eqnarray}
M_p = 2M_1\,\cos\frac{\alpha_{p}}{2}\,,
\end{eqnarray}
while the associated negative pole at $i(\pi-\alpha_p)$ represents the same
particle in the u-channel. In a non-unitary theory the s-channel bound state
may be represented by a negative pole (we will encounter this situation in
the theory \eqref{ift} at pure imaginary magnetic field $h$). Real poles
located in the mirror strip do not correspond to any stable particles;
borrowing terminology from the potential scattering, such poles are referred
to as the virtual states.

The (non)symmetry of the action \eqref{ift} generally allows the particle
$A_1$ to appear as the bound state pole in the $A_1 A_1$ scattering. This
"$\varphi^3$ property" means that the product in \eqref{poleform} always
\footnote{At special values of the parameters in \eqref{ift} this pole factor
can be canceled by another factor in the product; we will see that this
happens only at $h=0$.} involves the factor with
\begin{eqnarray}\label{alpha1}
\alpha_1 = \frac{2\pi}{3}\,.
\end{eqnarray}
This factor produces two poles in the PS; one at $\theta=2\pi i/3$,
interpreted as the s-channel pole associated with the particle $A_1$, and
another at $\theta=\pi i/3$, which is the corresponding u-channel pole.

Complex poles in the physical strip are forbidden by causuality. The complex
poles in the MS are generally interpreted as the resonance states. Because of
the crossing relation \eqref{crossing}, which can be equivalently written as
$S(-i\pi-\theta) = S(\theta)$, the pattern of the resonance poles in the MS
is symmetric with respect to the inversion ($=180^o$ rotation) around the
point $-i\pi/2$ in the $\theta$ plane. For this reason the resonance content
of the elastic $A_1 A_1$ scattering is completely determined by the poles
located in MS$^{(+)}$, Eq.\eqref{msp}. For a resonance pole at
$\theta=i\alpha_p\ \in MS^{(+)}$
\begin{eqnarray}
i\alpha_p = \beta_p - i\gamma_p
\end{eqnarray}
with real $\beta_p> 0\,$ and $\pi >\gamma_p>0\,$\, the s-channel complex mass
$M_p = 2M_1 \,\cos(\alpha_p/2)$ is
\begin{eqnarray}\label{mpres}
M_p = {\bar M}_p -i\Gamma_p\,, \qquad {\bar M}_p = 2M_1\,\cos\frac{\gamma_p}{2}\,\,\cosh\frac{\beta_p}{2}\,,
\quad \Gamma_p = 2M_1\,\sin\frac{\gamma_p}{2}\,\,\sinh\frac{\beta_p}{2}\,,
\end{eqnarray}
so that both ${\bar M}_p$ and $\Gamma_p$ are positive, and such poles admit
interpretation as the resonance states in the s-channel, with ${\bar M}_p$
and $\Gamma_p$ identified with the center of mass mean energy and the width
of the resonance state\footnote{Of course, only if $\gamma_p << \beta_p$ it
is fair to associate the pole at $i\alpha_p$ with a metastable state with a
long lifetime and well defined energy. However, for the luck of better term,
we refer to all the complex poles in MS$^{(+)}$ as the s-channel
resonances.}. Complex poles located in $MS^{(-)}$ are interpreted as the
u-channel resonances.

Because of the crossing symmetry \eqref{crossing} of the $A_1 A_1$
scattering, the resonance poles generally appear in the MS$^{(+)}$ in pairs:
$i\alpha_p$ comes along with $i(-\pi-\alpha_{p}^*)$. We call these pairs
"cross-conjugated"\footnote{Cross-conjugated points are complex conjugated in
terms of the variable $u=\theta+i\pi/2$. In fact, it is easy to see that as
the consequence of $S(i\alpha^*)=S^*(i\alpha)$ and \eqref{crossing} the
amplitude $S(u-i\pi/2)$ is real analytic function of $u$.}. Exceptions are
the poles with $\gamma_p=\pi/2$; for such poles $-\pi-\alpha_{p}^* =
\alpha_p\ \text{mod}\ 2\pi\,\mathbb{Z}$. We call such special poles self
cross-conjugated (not to be confused with real poles defined above).

\section{Pole evolution}

Let us return to the specific field theory \eqref{ift}. In this work we
concentrate attention at the high-T domain $m<0$, where the $h=0$ theory has
an unbroken symmetry $\sigma \to -\sigma$. As the result, at $h\neq 0$ the
most important characteristics (mass spectrum, S-matrix) are even functions
of $h$. Therefore in this domain it is convenient to consider the dependence
of the theory on the dimensionless variable
\begin{eqnarray}\label{xxidef}
\xi^2 = \frac{h^2}{(-m)^\frac{15}{4}}\,.
\end{eqnarray}
It is commonly assumed that, as the functions of this variable, the
characteristics of the theory admit analytic continuation to the whole
complex $\xi^2$-plane with the branch cut along the negative real axis, from
$-\infty$ to $-\xi_{0}^2$, as shown in Fig.3a. The numerical value
\begin{eqnarray}\label{xxi0}
\xi_{0}^2=0.035846(4)
\end{eqnarray}
\begin{figure}[ht]
\centering
\includegraphics[width=9cm]{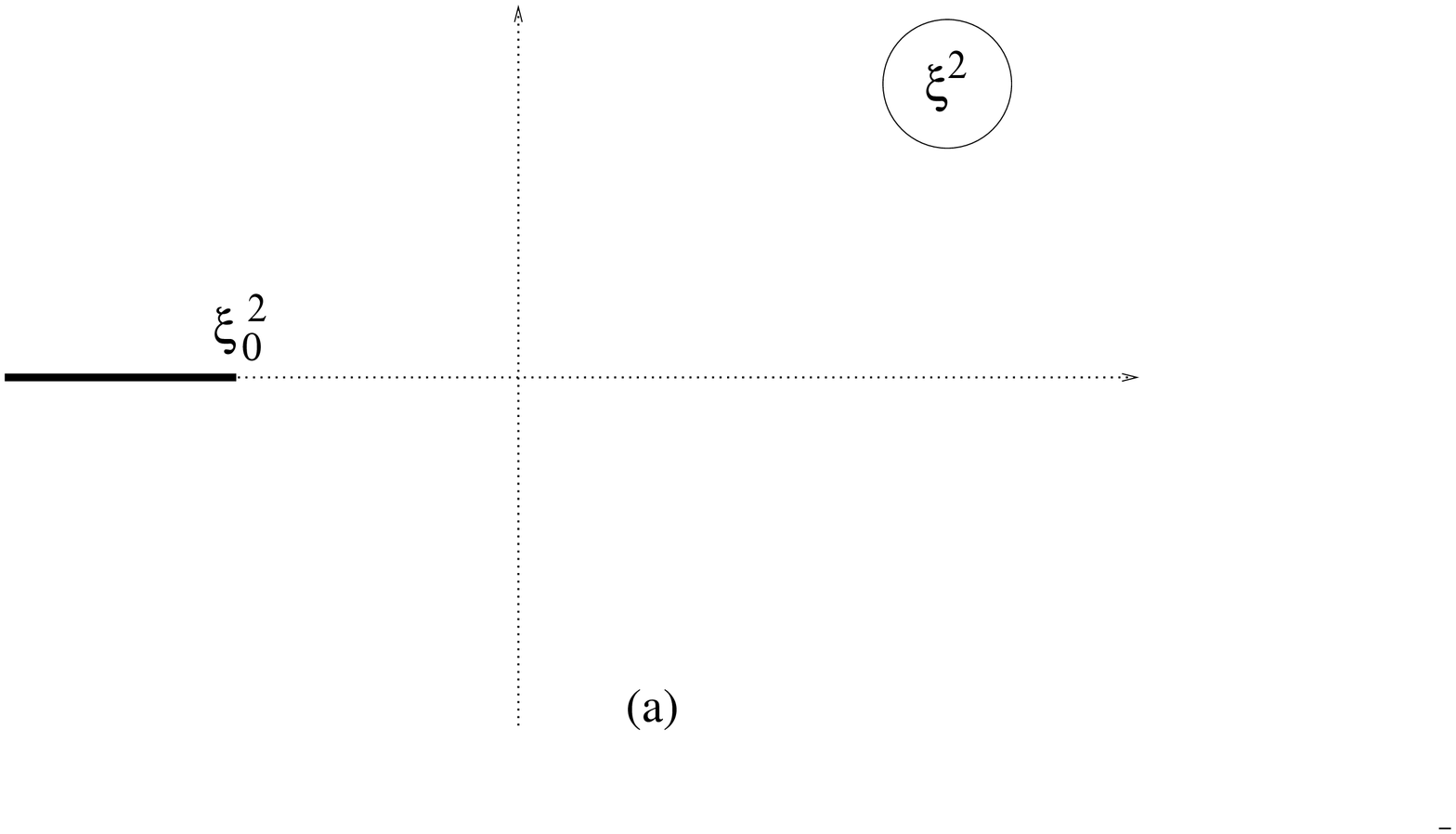}
\includegraphics[width=7cm]{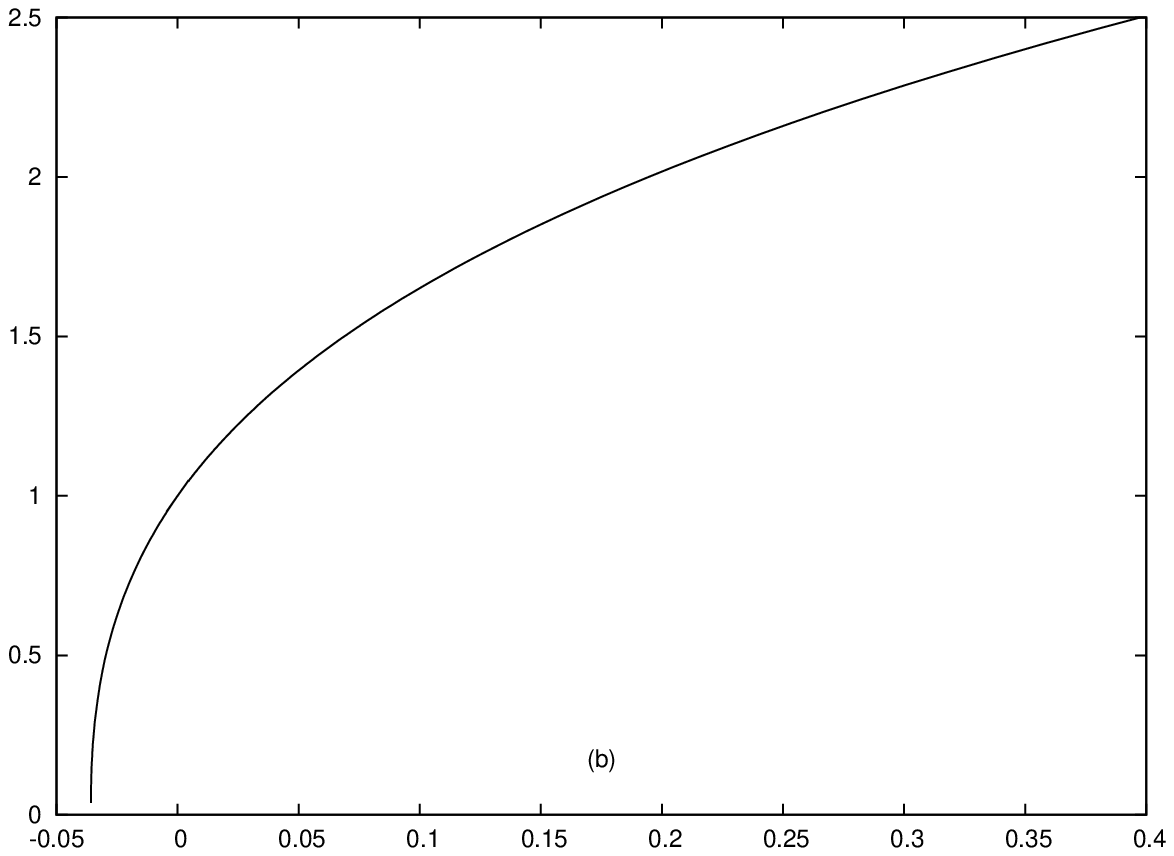}
\caption{\small{(a) Physical characteristics of IFT (e.g. S-matrix) are assumed
to admit analytic continuation onto the complex plane of $\xi^2$, with the
branch cut from $-\infty$ to the Yang-Lee singularity $-\xi_{0}^2$.
(b) Plot of the ratio $M_1/(-m)$ as the function of $\xi^2$, at real $\xi^2
> -\xi_{0}^2$ (The data is obtained via TFFSA \cite{zyl}). The mass turns to
zero at $-\xi_{0}^2$ according to \eqref{m1yl}. At large positive $\xi^2$ the
ratio $M_1/(-m)$ approaches asymptotic form
$M_{1}^{(0)}\,(\xi^2)^\frac{4}{15}$, where $M_{1}^{(0)}$ is given in
Appendix.}} \label{masses}\end{figure} was estimates in
\cite{fz1}\footnote{The estimate in \eqref{xxi0} is slightly better then the
one given in \cite{fz1}; it was obtained by detailed numerical analysis of
the mass $M_1(\xi^2)$ near the Yang-Lee critical point \cite{zyl}.}. The
point $-\xi_{0}^2$ is the so called Yang-Lee edge singularity. It is a
critical point in the sense that the mass $M_1$ of the particle $A_1$
vanishes at this point (see Fig.3b) \cite{Fisher}. The CFT associated with
this critical point was identified as the $\mathcal{M}_{2/5}$ nonunitary
minimal model with the central charge $c_{YL}=-22/5$ \cite{Cardy}.

At real positive $\xi^2$ the theory \eqref{ift} is unitary (both the
parameters $m$ and $h$ can be taken real). The negative part of the real axis
in Fig.3a is realized by pure imaginary values of the magnetic field $h$, so
here the unitarity is violated. However, at real $\xi^2 > -\xi_{0}^2$ the
theory remains "real" in the sense that the ground state energy and the
particle mass remains real, and the S-matrix still enjoys all the real
analyticity (but not positivity) conditions stated in Sect.2 \footnote{At
$\xi^2 < - \xi_0^2$ these properties are lost. The ground state becomes
two-fold degenerate, with the complex conjugate values of the associated
vacuum energies. The particle masses become complex as well, and all reality
conditions of the S-matrix described in Sect.2 are violated. We do not
discuss this interesting regime here.}. The most obvious manifestation of the
non-unitarity in this domain is negative residue of the $A_1$-particle pole
of $S(\theta)$ (see below).

In this work we discuss the theory \eqref{ift} at the values of the parameter
$\xi^2$ along the segment $\xi^2 > -\xi_{0}^2$ of the real axis in the
Fig.3a. The spectrum of stable particles is relatively simple - there is the
particle $A_1$ at all these values of $\xi^2$, and at sufficiently large
positive $\xi^2$ (at $\xi^2 > \xi_{2}^2$, and at $\xi^2 > \xi_{3}^2$,
respectively, where $\xi_2 = 0.253(5)$, $\xi_3 = 41(2)$) the stable particles
$A_2$ and $A_3$ appear in the spectrum \footnote{Also, at $\xi^2 = +\infty$,
which corresponds to one of the integrable points (see below), five
additional particles appear; at $\xi^2 < +\infty$ they become resonance
states, as we briefly discuss in Sect.4}. However, the spectrum of the
virtual and resonance states is more rich. Below I describe the evolution of
the pole structure of $S(\theta)$ as the parameter $\xi^2$ changes from
$+\infty$ to $0$, and then from $0$ to the Yang-Lee point $-\xi_{0}^2$. I
will pay most attention to the real poles; once a pole at $i\alpha_p$ leaves
the imaginary $\theta$-axis, I usually will not attempt to follow its fate as
a complex pole. The reasons for this are as follows. First, the low-energy
behavior of the the elastic scattering amplitude $S(\theta)$ is usually
dominated by its real-pole content. But more significantly, there still is
very little amount of data, both analytical and numeric, about high-energy
behavior of scattering amplitudes (see however \cite{ZZy}), and about high
energy resonance states in particular. The main goal of this note is to put
forward a scenario for the evolution of the real poles of $S(\theta)$ when
$\xi^2$ changes from $-\infty$ down to $-\xi_{0}^2$. However, I will make
remarks about evolution of some resonance poles in Sect.4.

Although the parameter $\xi^2$ is quite suitable for the needs of the present
discussion, to facilitate comparison with Refs.\cite{fz1,fz3} I will
frequently discuss simultaneously in terms of the related scaling parameter
$\eta$ defined in Eq.\eqref{etadef}. Generally, these parameters are related
by the complex-analytic map
\begin{eqnarray}
\xi^2 = \frac{1}{(-\eta)^\frac{15}{4}}\,, \qquad
\left|\text{arg}(-\eta)\right| \leq \frac{4\pi}{15}\,,
\end{eqnarray}
with the branch chosen in such a way that the negative part of the
$\eta$-axis is mapped to real positive $\xi^2$. Then, while positive real
$\xi^2$ are represented by negative real $\eta$, negative $\xi^2$ correspond
to complex $\eta$ along certain rays in the $\eta$-plane,
\begin{eqnarray}\label{yrays}
\xi^2 <0 \ : \quad \rightarrow\quad  \eta = y \,e^{\pm\frac{4\pi i}{15}}\quad
\text{with real} \ \ y<0\,.
\end{eqnarray}
The segment $-\xi_{0}^2 < \xi^2 < 0$ is given by $y < -Y_0$, where
\begin{eqnarray}
Y_0 =  (\xi_{0}^2)^\frac{4}{15} =  2.42929(2)\,.
\end{eqnarray}

\subsection{Integrable points}

There are three special points in the parameter space of the theory
\eqref{ift} where the amplitude $S(\theta)$ is known exactly:
\bigskip

\textbf{(a)}: $\xi^2=+\infty$ ($\eta=0$), which corresponds to $m=0$, $h\neq
0$. At this point the theory \eqref{ift} is integrable, and its spectrum
involves 8 stable particles $A_1, \ A_1,\ ...,\ A_8$ whose masses $M_1,\
M_2,\ ..., M_8$ are given by the components of the Perron-Frobenius vector of
the Cartan matrix of the Lie algebra $E_8$. The scattering theory is
described by the reflection-less factorizable S-matrix \cite{e8}. $A_1$ is
the lightest particle, and $A_1 A_1$ elastic scattering is given by
\eqref{poleform} with $\Delta(\theta)=0$ and three pole factors,
\begin{eqnarray}\label{szero}
S(\theta) = \frac{\sinh\theta+i\sin(2\pi/3)}{\sinh\theta-i\sin(2\pi/3)}
\,\frac{\sinh\theta+i\sin(2\pi/5)}{\sinh\theta-i\sin(2\pi/5)}
\,\frac{\sinh\theta+i\sin(\pi/15)}{\sinh\theta-i\sin(\pi/15)}\,.
\end{eqnarray}
All the poles here are located in the physical strip, at $\theta= i\alpha_p$
and $\theta=i(\pi-\alpha_p)$, $p=1,2,3$, where
\begin{eqnarray}
\alpha_1 =\frac{2\pi}{3}\,,\qquad\alpha_2 = \frac{2\pi}{5}\,,
\qquad \alpha_3 = \frac{\pi}{15}\,.
\end{eqnarray}
While the pole at $\theta = 2\pi i/3$ represents the s-channel particle $A_1$
itself, the poles at $\theta=2\pi i/5$ and $\theta=i\pi/15$ correspond to
heavier stable particles $A_2$ and $A_3$, with the masses
\begin{eqnarray}
M_2 = 2 M_1\,\cos(\pi/5)\,, \qquad M_3 = 2 M_1\,\cos(\pi/30)\,.
\end{eqnarray}
As usual, the poles at $i(\pi-\alpha_p)$ are the u-channel manifestations of
the same particles. The pattern of poles and zeroes of $S(\theta)$ is shown
in Fig.4a. Although the poles associated with the higher stable particles
$A_4,\ ...\,, A_8$ are not present in $S(\theta)$, such poles show up in
other elastic scattering amplitudes \cite{e8}. Observe that (i) $S(\theta)$
has no zeros in the PS, and (ii) There is one u-channel pole
$i(\pi-\alpha_p)$ in between any two s-channel poles $i\alpha_i$. These two
conditions, together with easily verified $S(0) = -1$, guarantee that the
residues $r_p$ at the s-channel poles $i\alpha_p$, $p=1,2,3$, are all
positive.
\bigskip

\textbf{(b)}: $\xi^2 =0$ ($\eta=-\infty$), which corresponds to $h=0$ with $m
<0$. At zero magnetic field the \eqref{ift} reduces to the theory of free
Majorana fermi field with the mass $m$. The free particle is $A_1$, hence at
this point $M_1 =m$, and
\begin{eqnarray}\label{sh0}
S(\theta) = -1\,.
\end{eqnarray}

\bigskip

\textbf{(c)}: $\xi^2 = -\xi_{0}^2+0$\ \ ($\eta=(-Y_0-0)\,e^{\pm\frac{4\pi
i}{15}}$). The integrable theory appears in the scaling limit $\xi^2 \to
-\xi_{0}^2$ near the Yang-Lee critical point. As was already mentioned, as
$\xi^2$ approaches the Yang-Lee point the mass $M_1$ of the particle $M_1$
vanishes as
\begin{eqnarray}\label{m1yl}
M_1 \sim (\xi^2 + \xi_0^2)^{5/12}\,,
\end{eqnarray}
while the masses of all resonances (there are no other stable particles near
the Yang-Lee point) remain finite in this limit. Therefore, when measured in
the units of $M_1$, all these masses depart to infinity at $\xi^2 \to \xi_0^2
+ 0$. The resulting scattering theory of the particles $A_1$ turns out to be
integrable, with the $2\to 2$ S-matrix given by \cite{cardy-mussardo}
\begin{eqnarray}\label{syl}
S(\theta) = \frac{\sinh\theta+i\sin\alpha_1}{\sinh\theta-i\sin\alpha_1}\,, \qquad
\alpha_1 = \frac{2\pi}{3}\,.
\end{eqnarray}
Except for the poles at $i\alpha_1$, Eq.\eqref{alpha1}, and
$i(\pi-\alpha_1)$, which correspond to the s- and u- channel manifestations
of the $A_1$, there are no real or complex poles. Note that $S(0)=-1$, and
that $r_1 = -2\sqrt{3}$ is negative.

\subsection{Positive $\boldsymbol{\xi^2}$ (real negative $\boldsymbol{\eta}$)}

The goal of this subsection is to analyze how the pattern of poles in
\eqref{szero} (see Fig.4a) evolves into the trivial pattern (no poles, no
zeros) associated with the free particle S-matrix \eqref{sh0} when $\xi^2$
changes from $+\infty$ down to $0$. The arguments will be based partly on the
general properties of the S-matrix (see Sect.2), and partly on the numerical
and perturbative data about the mass spectrum of \eqref{ift}. Since the
theory \eqref{ift} with real $m$ and $h$ is unitary, the general properties
include the condition that all s-channel poles associated with stable
particles remain positive during the evolution. We also assume that the
parameters
\begin{eqnarray}
B_p = \sin\alpha_p
\end{eqnarray}
depend on $\xi^2$ in an analytic way along the whole segment
$(-\xi_{0}^2:+\infty]$ \footnote{Generally, there are no good reasons to
expect $B_p$ themselves to be analytic in the whole range of the coupling
parameter. Rather, one should expect analyticity of the coefficients of the
polynomial $\prod (\sinh\theta -iB_p)$, while $B_p$ are allowed to have
algebraic singularities. We will see below that this subtlety is irrelevant
in the simple scenario suggested below.}. The pole at $\theta = i\alpha_1$ is
interpreted as the s-channel particle $A_1$, and therefore $\alpha_1$ remains
fixed to the value \eqref{alpha1} during the whole evolution. We need to
figure out the behavior of the remaining $\alpha_p$.

When $\xi^2$ is large (equivalently, $\eta$ is small) the theory \eqref{ift}
is in the perturbative domain around the integrable point {\bf (a)}. The
leading corrections to the mass ratios $M_2/M_1$ and $M_3/M_1$ were found in
\cite{Mussardo1,delfino2}, using the form factor perturbation theory. From
these, and the numerical data, one finds
\begin{eqnarray}
\alpha_2 = \frac{2\pi}{5} + \alpha_2^{(1)}\,\eta + \alpha_{2}^{(2)}\,\eta^2 +
...\,,\qquad
\alpha_3 = \frac{\pi}{15} + \alpha_3^{(1)}\,\eta + \alpha_{3}^{(2)}\,\eta^2 +
...\,,
\end{eqnarray}
\begin{eqnarray}
&&\alpha_{2}^{(1)} = 0.378325... \,, \qquad \alpha_2^{(2)} = -0.1153\,, \\
&&\alpha_{3}^{(1)} = 1.35226...\,, \qquad \ \, \alpha_3^{(2)} = -1.10\,,
\end{eqnarray}
When $\eta$ decreases from $0$ to negative values, the parameters $\alpha_2$
and $\alpha_3$ both decrease, so that the associated poles move towards zero
(see Fig.4b). Simultaneously, a number of additional complex poles appear;
these resonance poles are ignored in Fig's 4 and 5. I postpone comments about
them to Sect.4. At certain value $\eta=\eta_3$ the pole at $i\alpha_3$
crosses zero. At this point $M_3/M_1=2$. Numerical data yields
\begin{eqnarray}\label{eta3}
\eta_3 = - 0.138(6) \qquad\quad (\xi_{3} = 41(3))
\end{eqnarray}

\begin{figure}[ht]
\centering
\includegraphics[width=8cm]{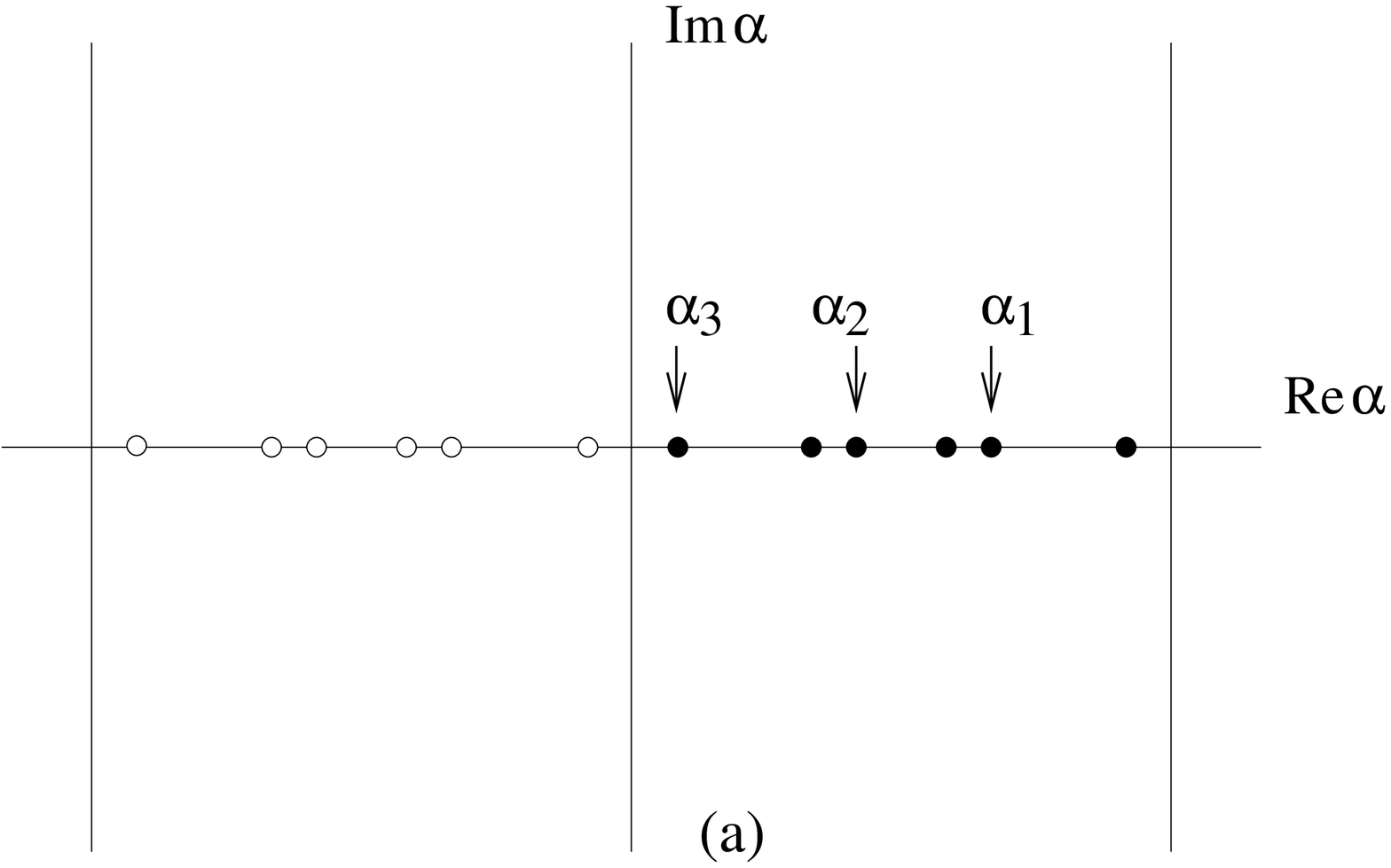}
\includegraphics[width=8cm]{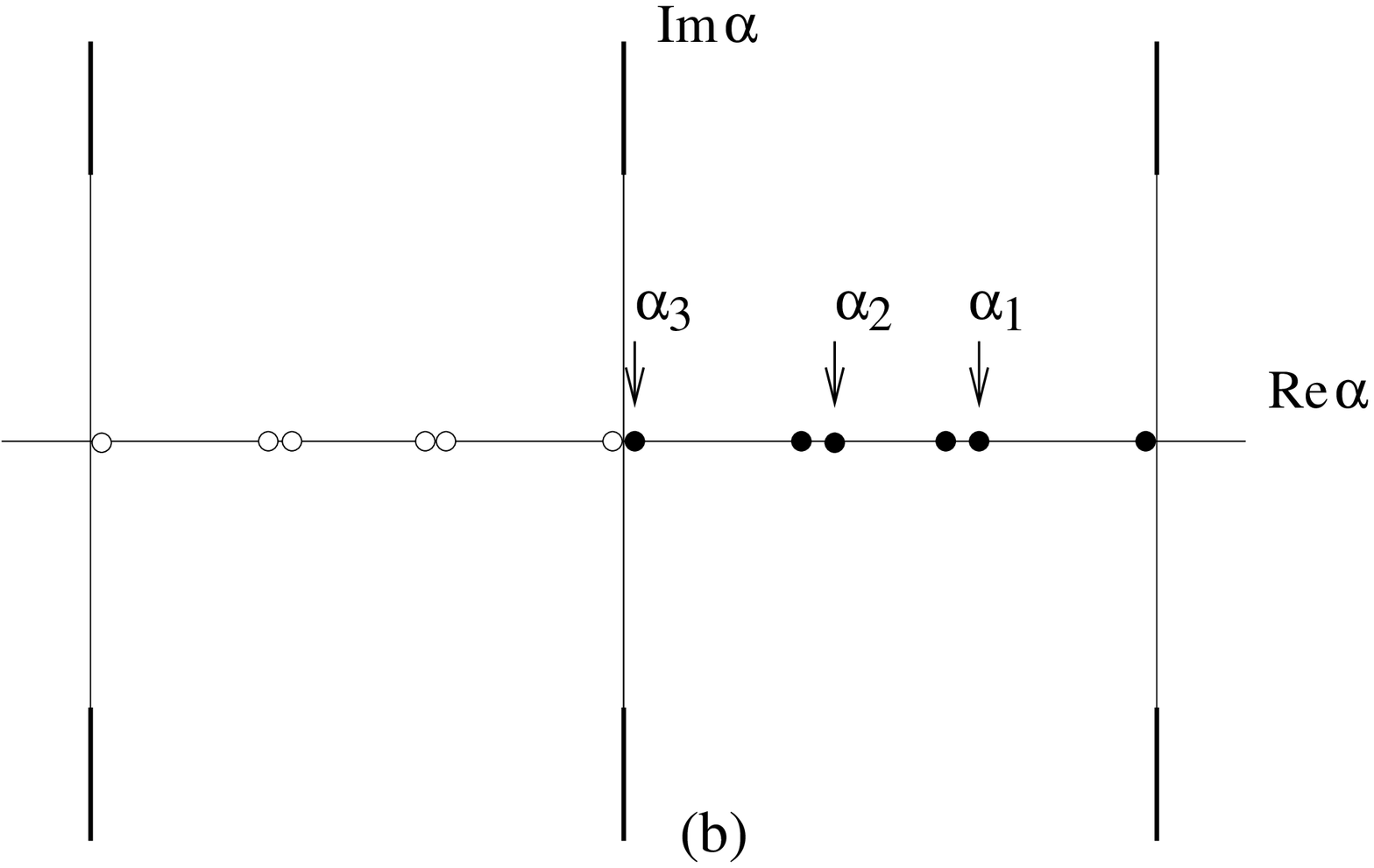}
\vskip 0.3cm
\includegraphics[width=8cm]{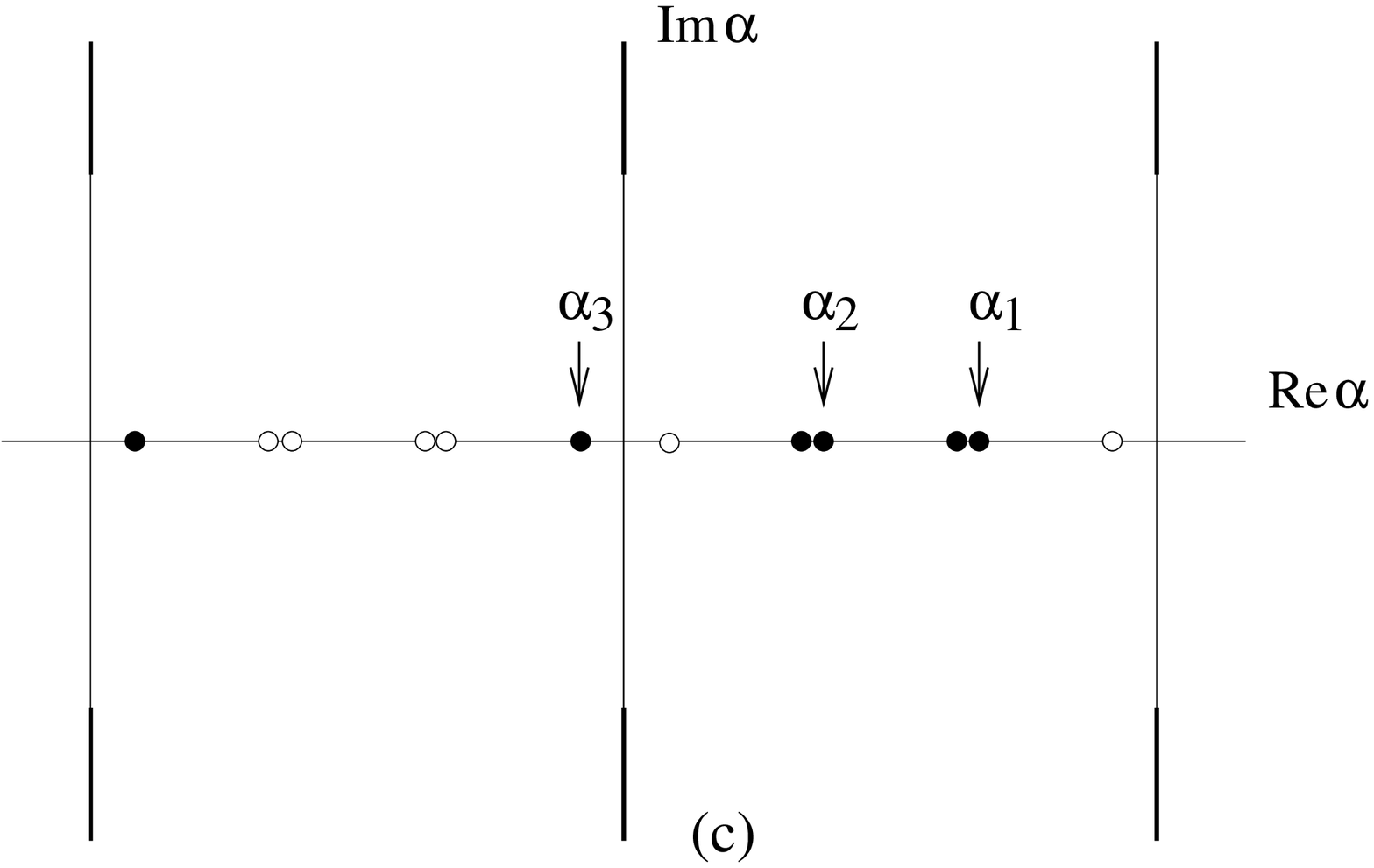}
\includegraphics[width=8cm]{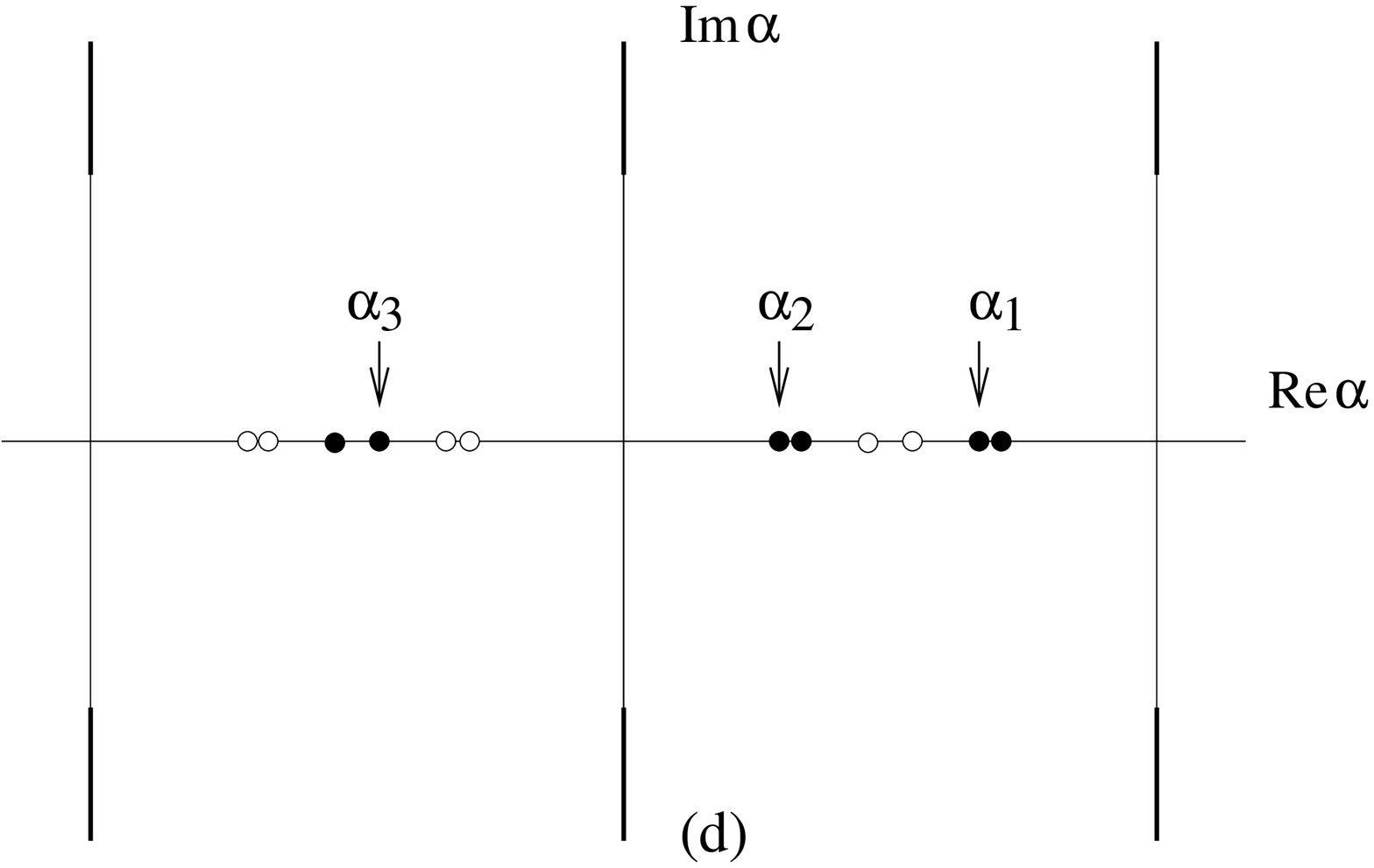}
\caption{\small{Real poles $\bullet$ and zeros $\circ$ of $S(i\alpha)$ in the
complex plane of the variable $\alpha=-i\theta$, at some values of $\eta$.
(a) Integrable point $\eta=0.00$. The poles and zeros of \eqref{szero} are shown
 (b) $\eta=-0.08$. The poles $\alpha_2$ and $\alpha_3$ move towards zero, but
 are still in the PS. At nonzero $\eta$ a number of complex poles
 associated with the resonances $A_4$, $A_5$, ... also appear (see Sect.4);
 in this and the subsequent drawings I ignore such poles. (c) $\eta=-0.27$. The
 pole $\alpha_3$ has crossed into the MS, the pole $\alpha_2$ has moved further
 towards zero. (d)  $\eta=-0.49$. The pole $\alpha_2$ interchanges order with
 the pole $\pi-\alpha_1=\pi/3$. Simultaneously, the mirror zero at $-\alpha_3$
 moves over the same point $\pi/3$.}}
 \label{patterns}
\end{figure}

\noindent and
\begin{eqnarray}
\alpha_3(\eta) = \alpha_{3}'\,(\eta-\eta_3) + O\left((\eta-\eta_3)^2\right)\,,
\qquad \alpha_{3}' \approx 1.65\,.
\end{eqnarray}
At $\eta < \eta_3$ the pole $i\alpha_3$ leaves the PS and enters the MS
(Fig.4c), so that the particle $A_3$ disappears from the spectrum of stable
particles, becoming a virtual state. After crossing to MS $i\alpha_3$ becomes
a negative pole (the residue $r_p \sim \sin\alpha_p$). However, we will still
associate the pole at $i\alpha_3$ with the "particle" $A_3$, now a virtual
state, and call it the "$A_3$ pole". Note that although below $\eta_3$ the
number of stable particles in \eqref{ift} changes from three to two, the
number of pairs of real poles in $S(\theta)$ remains equal to three all the
way down to somewhat lower value of $\eta$ (see Eq.\eqref{eta33} below).

When $\eta$ continues to decrease from $\eta_3$ down, the next interesting
event occurs at
\begin{eqnarray}
\eta_{12} = -0.477(4) \qquad\quad (\xi_{12} = 4.01(6))
\end{eqnarray}
where the mass ratio $M_2/M_1$ reaches the value $\sqrt{3}$. At this point
the pole at $i\alpha_2$, on its way towards zero, crosses the point
$i(\pi-\alpha_1)=i\pi/3$ where the u-channel pole of the particle $A_1$ sits.
After that, these poles interchange their order along the imaginary
$\theta$-axis. Without additional poles or zeros around, after such "pole
crossing" the formerly positive pole $i\alpha_2$ would become a negative one,
while the pole at $i\pi/3$ would become a positive pole. But unitarity
demands that the poles must retain their signatures as long as $\eta$ remains
real. The only way this can happen is if there is a zero of $S(\theta)$
located in the PS which passes through $i\pi/3$ at the same point
$\eta=\eta_{12}$. There is only one pair of zeros in the PS in this domain of
$\eta$ - the zeros at $-i\alpha_3$ and $i(\pi+\alpha_3)$ (these are the
"mirror images" of the poles $i\alpha_3$ and $i(-\pi-\alpha_3)$ which at
$\eta < \eta_3$ are already in the MS). Therefore, we must have
\begin{eqnarray}\label{eta12a}
\eta_{12}: \qquad \alpha_2=\pi/3\,, \quad \alpha_3=-\pi/3\,.
\end{eqnarray}
The pattern of poles at $\eta$ immediately above and immediately below
$\eta_{12}$ are shown in Fig.4c and Fig.4d.

As $\eta$ decreases further from $\eta_{12}$, already negative
$B_3=\sin\alpha_3$ continues to decrease, and at certain $\eta=\eta_{33}$ it
crosses $-1$.  At this point the poles $i\alpha_{3}$ and $-i(\pi+\alpha_3)$
collide at the middle of MS,
\begin{eqnarray}\label{eta33a}
\eta_{33}:  \qquad \alpha_{3} = -(\pi+\alpha_3) = -\pi/2\,,
\end{eqnarray}

\begin{figure}[ht]
\centering
\includegraphics[width=8cm]{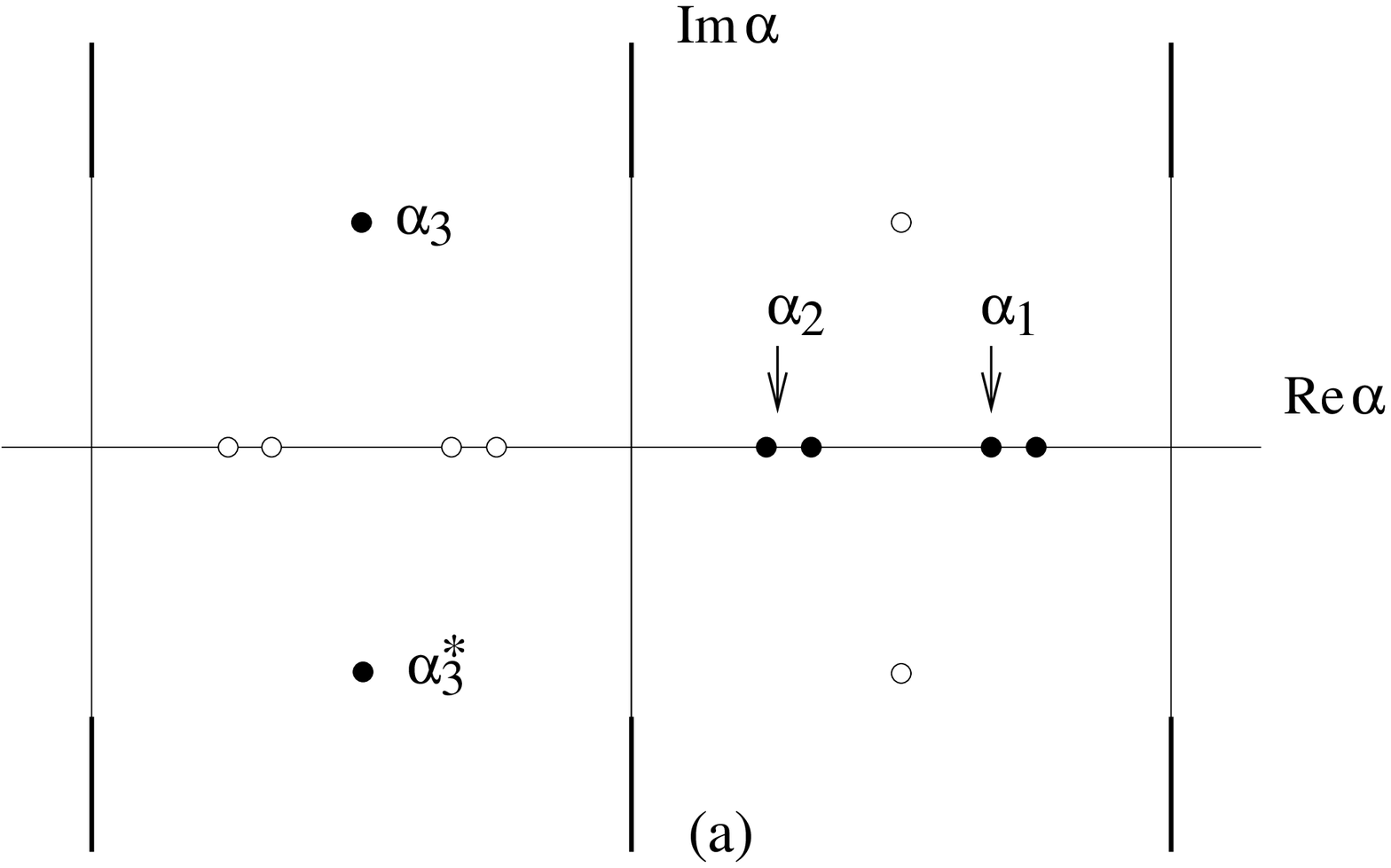}
\includegraphics[width=8cm]{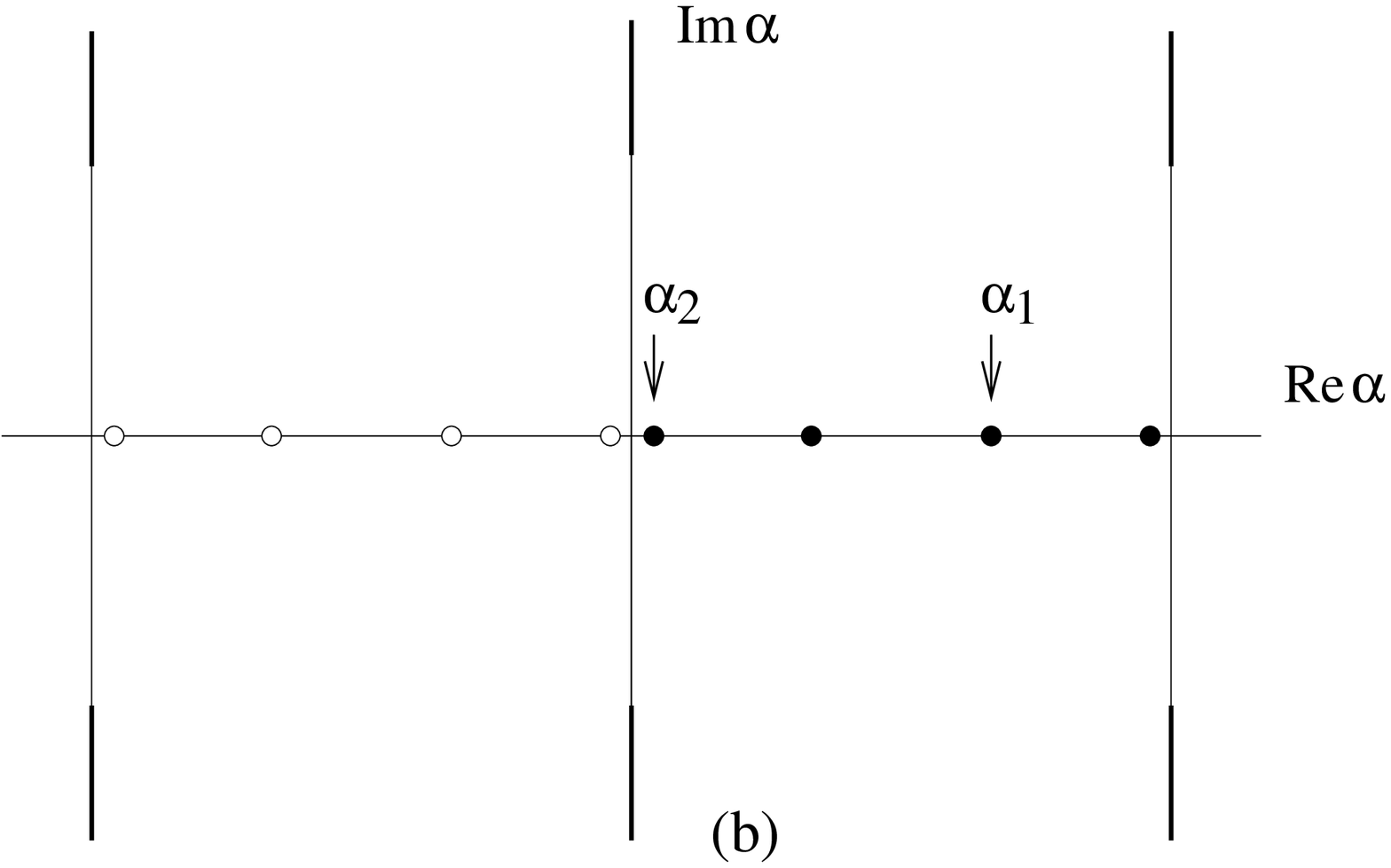}
\vskip 0.3cm
\includegraphics[width=8cm]{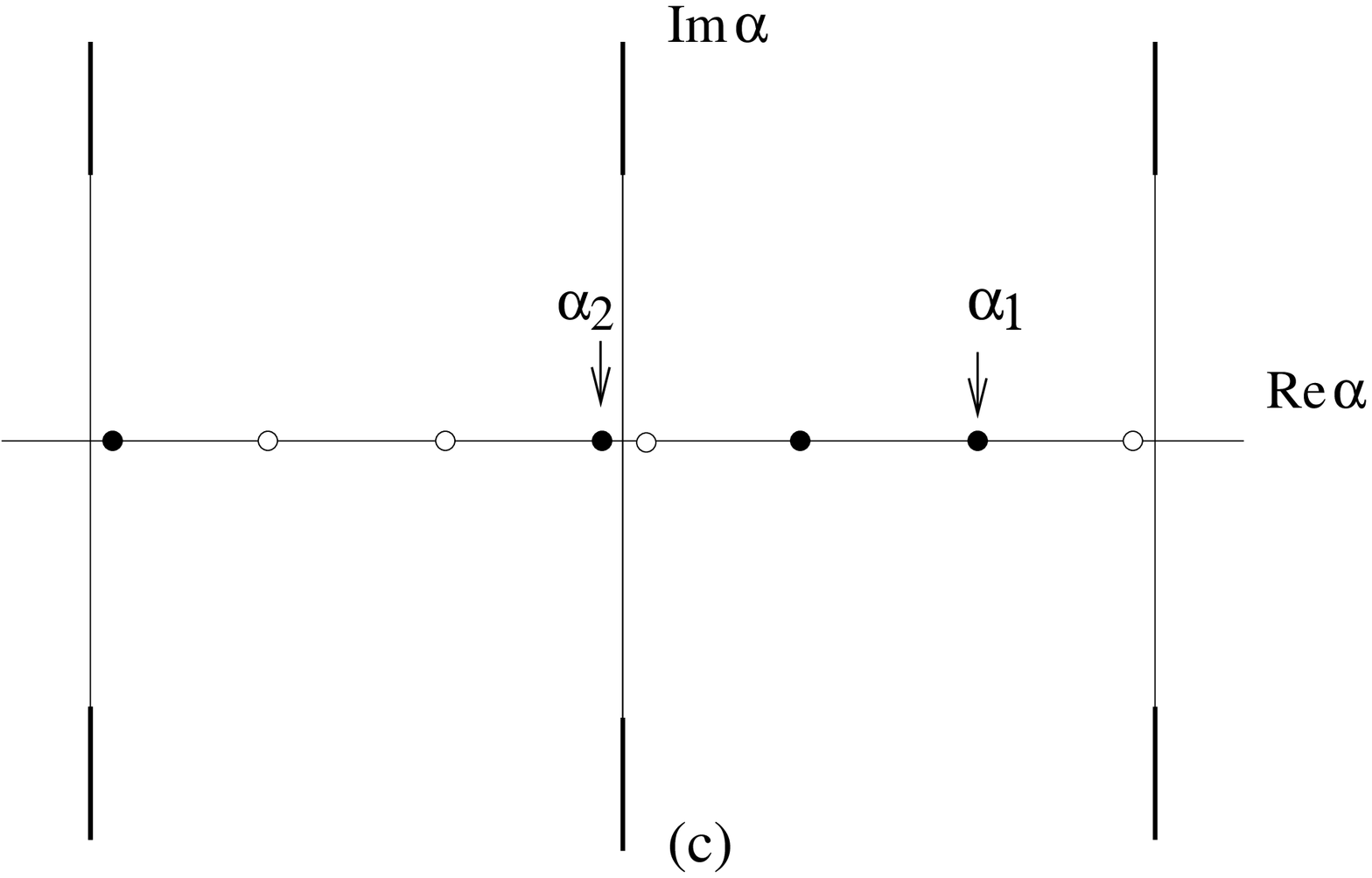}
\includegraphics[width=8cm]{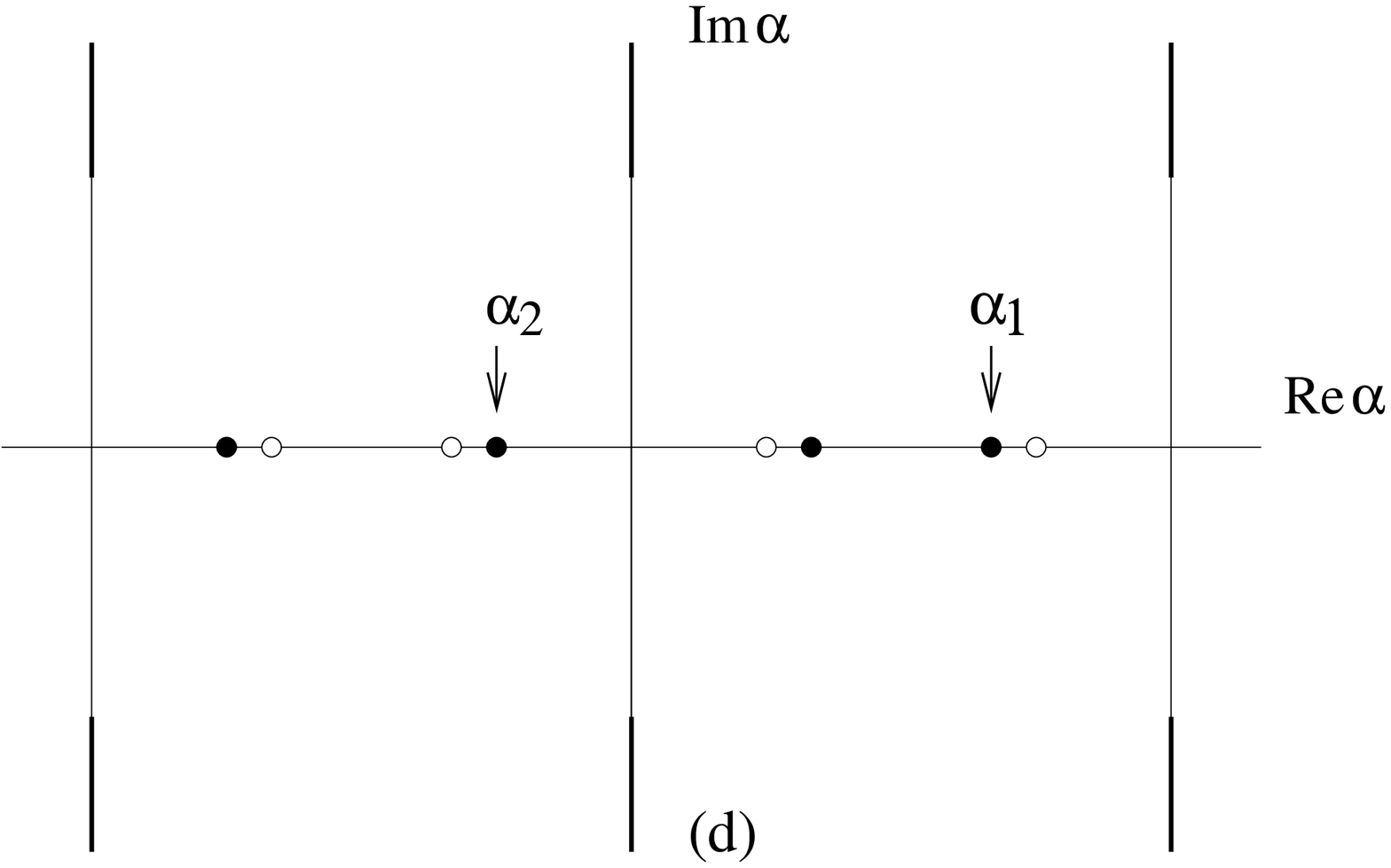}
\caption{\small{Poles $\bullet$ and zeros $\circ$ of $S(i\alpha)$ in the
complex $\alpha$-plane, at some values of $\eta$. Except for the pair
$\alpha_3, \alpha_{3}^*$, only real poles are shown. (a) $\eta = -0.94$.
After colliding at $-\pi/2$ (which happens at $\eta_{33}$,
Eq.\eqref{eta33}) the poles at $\alpha_3$ and $-\pi-\alpha_3$ move
away from the real $\alpha$-axis. (b) $\eta=-1.87$. The pole at $\alpha_2$
gets closer to zero (c) $\eta=-2.29$. The pole at $\alpha_2$ has left the
PS. The only stable particle left is $A_1$.
(d)$\eta=-4.35$. The pole at $\alpha_2$ approaches the fixed zero at $-\pi/3$,
and its residue becomes small, Eq.\eqref{r1alpha}. In the limit $\eta
\to -\infty$ the zero cancels the pole, resulting in \eqref{sh0}.}
 }\label{patterns}
\end{figure}

\noindent and at $\eta<\eta_3$ they become a pair of self cross-conjugated
complex poles, moving away from the imaginary $\theta$-axis as $ -i\pi/2\mp
\beta_3$ with real $\beta_3$ (Fig.5a). Therefore, below $\eta_{33}$ the
number of real poles changes from three to two. Note that at $\eta_{33}$ the
analytic continuation of $M_3/M_1$ takes the value $\sqrt{2}$, and
immediately below this point $M_3$ becomes complex. The numerical value
\begin{eqnarray}\label{eta33}
\eta_{33} = -0.51(2) \qquad\quad (\xi_{33}=3.5(3))
\end{eqnarray}
\begin{figure}[ht]
\centering
\includegraphics[width=10cm]{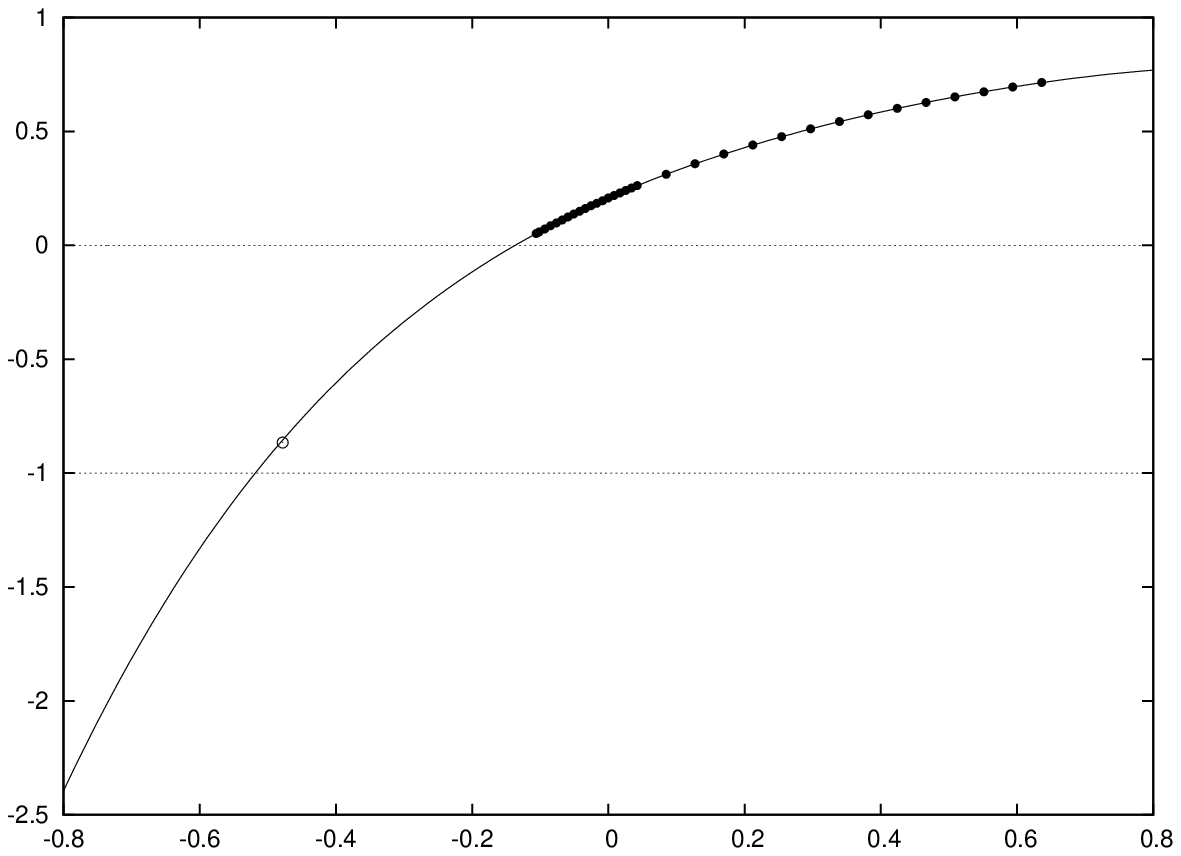}
\caption{\small{Behavior of $B_3 = \sin\alpha_3$ as the function of real
$\eta$ near $\eta=0$. The bullets $\bullet$ show the result of analysis
of direct TFFSA data in the domain $\eta > \eta_3$, where $A_3$ is visible
as the stable particle. The solid line is the polynomial
extrapolation of this data. The circle $\circ$ marks the value
$B_3(\eta_{12})=-\sqrt{3}/2$, see Eq.\eqref{eta12a}; its position near
the solid line indicates the quality of the extrapolation. The graph of
the function $B_3(\eta)$ crosses $0$ at $\eta_3$, Eq.\eqref{eta3}, and $-1$
at $\eta_{33}$, Eq.\eqref{eta33}.}}\label{patterns}
\end{figure}
\noindent was estimated by extrapolating the data for $B_3$ from the domain
of $\eta$ above $\eta_3$, where the mass ratio $M_3/M_1$ is directly
available from TFFSA. The plot in Fig.6 suggests that as $\eta$ decreases
further down from $\eta_{33}$ the poles $i\alpha_3$ and $i\alpha_{3}^*$
rapidly move away from the real axis, so that almost immediately below
$\eta_{33}$ the $A_3$ pole becomes a high energy resonance. At the moment we
do not have anything but speculations about its fate at $\eta$ substantially
below $\eta_{33}$ \footnote{Simple form \eqref{sh0} suggests that this pair
of poles must depart to infinity either at $\xi^2=0$ or at some intermediate
value between $0$ and $\xi_{33}^2$. The result of Ref.\cite{ZZy} seems to
rule out the first possibility.}.

As for the remaining real poles, at $\eta$ below $\eta_{33}$ the pole
$i\alpha_2$ continues to move towards zero, and at $\eta = \eta_2$ it too
crosses into MP. Again, from TFFSA data on $M_2/M_1$ we have
\begin{eqnarray}\label{eta2}
\eta_2 = -2.08(2)\qquad \quad (\xi_{2} =0.253(5))\,.
\end{eqnarray}
Below $\eta_2$ the particle $A_2$ ceases to exist as a stable particle,
becoming a virtual state instead, but as with $i\alpha_3$, we will continue
to call $i\alpha_2$ the "$A_2$ pole". At $\eta < \eta_2$ the real spectrum of
stable particles of \eqref{ift} involves a single neutral particle $A_1$. The
patterns of poles and zeros of $S(\theta)$ just above and just below $\eta_2$
are shown in Figures 5b and 5c, respectively. Note that since below $\eta_2$
the $A_2$ pole remains real, the analytic continuation of $M_2$ remains real
and below $2 M_1$ as well; in fact, it is real for all $\xi^2 > \xi_{22}^2$
(see below).

As $\eta$ decreases further, the pole $i\alpha_2$ sinks deeper into MS, and
eventually it approaches the point $-i\pi/3$ where the fixed zero (mirror
image of the u-channel $A_1$ pole $i(\pi-\alpha_1)=i\pi/3$) sits. When
$\alpha_2$ is close to $-\pi/3$ the residue $r_1$ at the $A_1$ pole
$i\alpha_1=2\pi i/3$ becomes small (recall that in this limit the mirror zero
of the u-cannel $A_2$ pole $i(\pi-\alpha_2)$ approaches the $A_1$ pole, as
seen in Fig.5d),
\begin{eqnarray}\label{r1alpha}
r_1\ \sim\  \alpha_2+\pi/3\,.
\end{eqnarray}
This is exactly what is expected in \eqref{ift} at small $h$. Indeed, the
residue $r_1$ vanishes at $h=0$ since we have to have \eqref{sh0}, and the
leading perturbative contribution is \cite{ZZy}
\begin{eqnarray}\label{r1pert}
r_1 = 36\,{\bar s}^2\,\xi^2 + O(\xi^4)\,,
\end{eqnarray}
where ${\bar s} = 2^{1/{12}}\,\,e^{-{3\over 2}\,\zeta'(-1)} =
1.35783834170660...$ is the constant appearing in the one-particle matrix
element $\langle \,0\mid \sigma(0)\mid A_1 \,\rangle = {\bar
s}\,\,(-m)^\frac{1}{8}$ \cite{spinspin}. We conclude that at large negative
$\eta$ $\ \alpha_2$ approaches $-\pi/3$ from above as
\begin{eqnarray}
\alpha_2 = - \frac{\pi}{3} + \frac{36\,{\bar s}^2}
{(-\eta)^\frac{15}{4}}
+ O\left((-\eta)^{-\frac{15}{2}}\right)\,.
\end{eqnarray}
In the limit $\eta=-\infty$ (i.e. at $\xi^2=0$) the pole $i\alpha_2$ gets
canceled by the fixed zero at $-i\pi/3$, and thus all real poles disappear,
consistently with \eqref{sh0}. Of course, \eqref{sh0} also demands that all
the resonance poles (which we left without attention so far) disappear in
this limit as well.

\subsection{Negative $\boldsymbol{\xi^2 > \xi_{0}^2}$}

In the previous subsection we have described the evolution of the real poles
of $S(\theta)$ when $\xi^2$ changed from $+\infty$ to $0$ (equivalently, real
$\eta$ changed from $0$ to $-\infty$). This evolution can be extended to
negative $\xi^2$ above the Yang-Lee point $-\xi_{0}^2$. The latter values
correspond to pure imaginary $h$ between zero and
$\pm\,i\xi_{0}\,|m|^{15/8}$. In terms of $\eta$ we will be dealing with the
values along the rays \eqref{yrays} with $-\infty<y<-Y_0$.

As we have seen above, at small positive $\xi^2$ the $A_2$ pole is located
close to the right of the fixed zero $i\pi/3$, and at $\xi^2=0$ this zero
exactly cancels the pole, leading to the trivial amplitude \eqref{sh0}. As
$\xi^2$ becomes small negative the $A_2$ pole re-emerges to the left of the
fixed zero $i\pi/3$. In terms of the variable $y$ defined in Eq.\eqref{yrays}
we have
\begin{eqnarray}
\alpha_2 = - \frac{\pi}{3} - \frac{36\,{\bar s}^2}
{(-y)^\frac{15}{4}}
+ O\left((-y)^{-\frac{15}{2}}\right)\,.
\end{eqnarray}
\begin{figure}[ht]
\centering
\includegraphics[width=8cm]{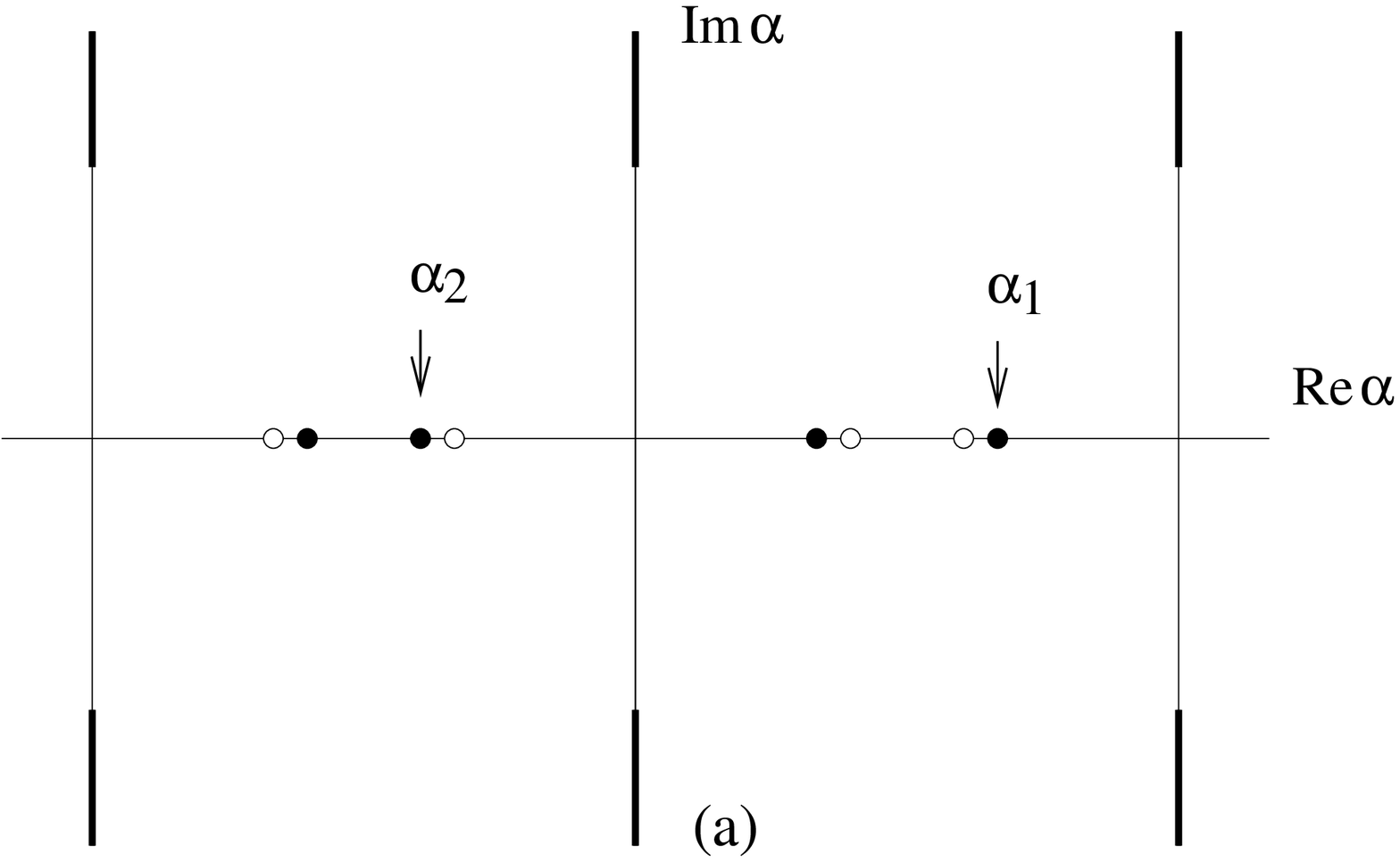}
\includegraphics[width=8cm]{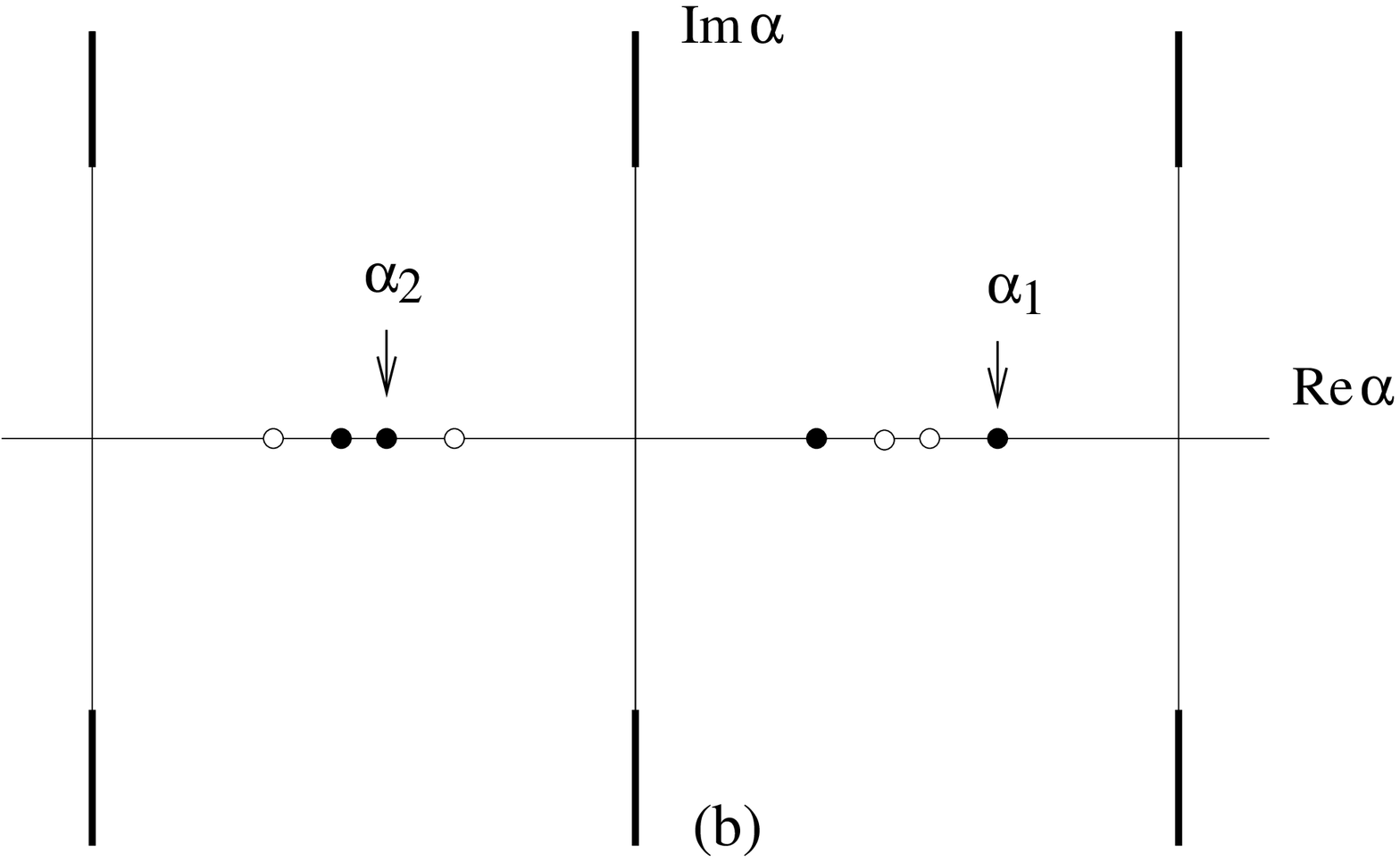}
\vskip 0.3cm
\includegraphics[width=8cm]{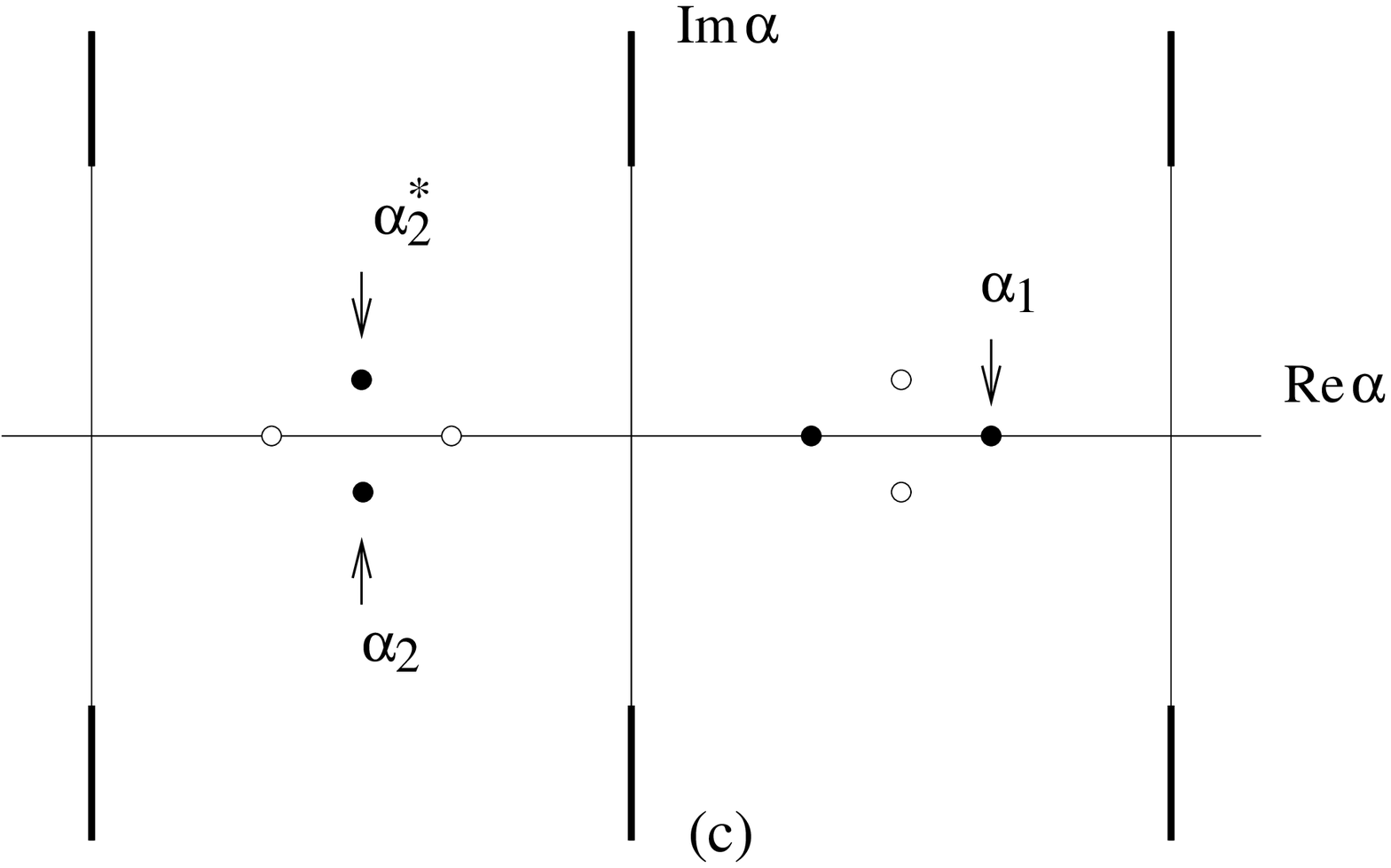}
\includegraphics[width=8cm]{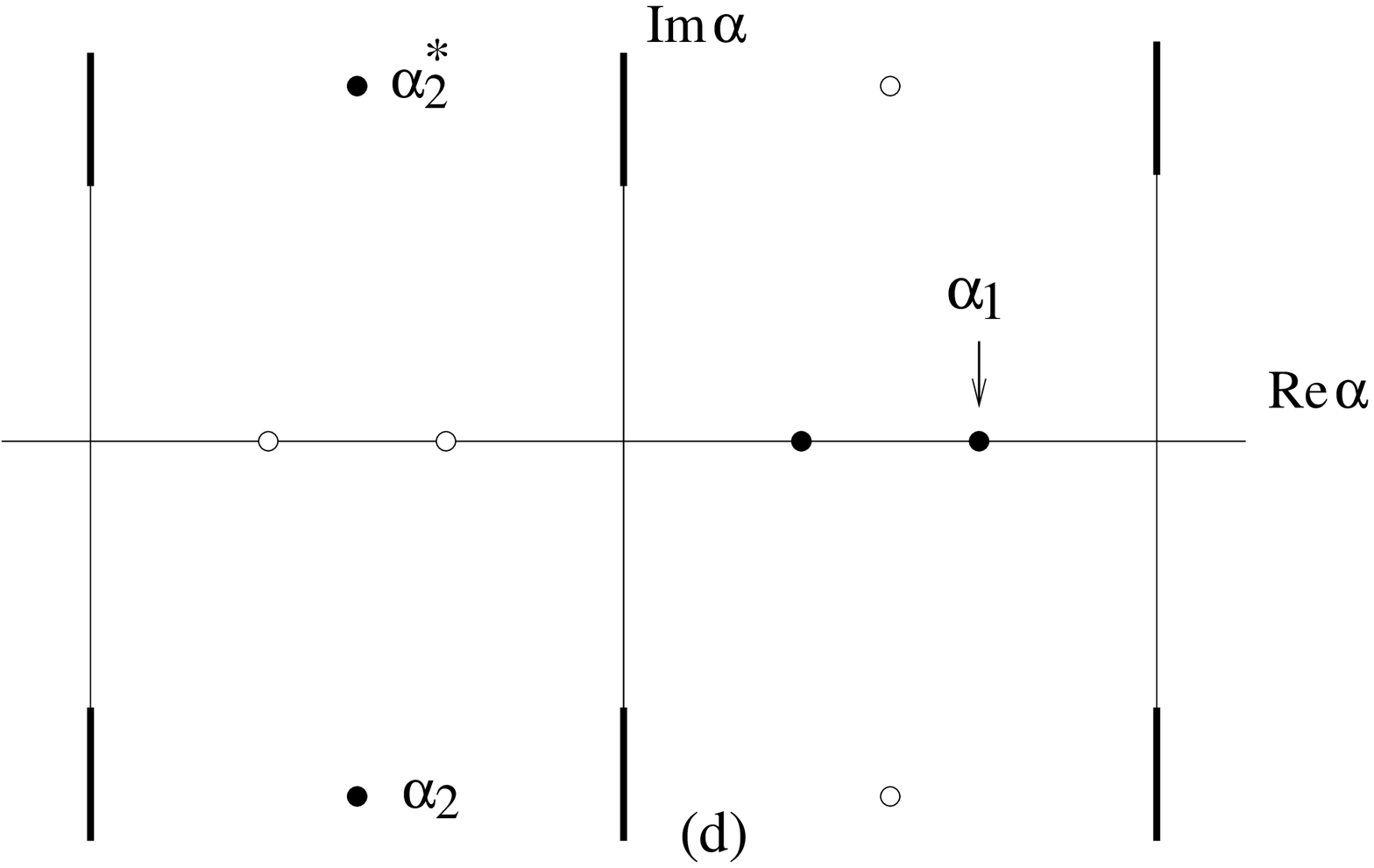}
\caption{Poles and zeros of $S(i\alpha)$ in the complex
$\alpha$-plane, at some values of the scaling parameter
corresponding to pure imaginary magnetic field $h$, i.e. with $\eta$ is taken
along the rays \eqref{yrays}.
(a) $\ y=-5.10$. The $A_2$ pole in the MS moves away to the left
from the zero at $-i\pi/3$. (b) $y=-4.75$. The real poles at
$\alpha_2$ and $-\pi-\alpha_2$ approach each other. (c) $y=-4.00$. After
colliding at $-\pi/2$ the poles at $\alpha_2,-\pi-\alpha_2$ become complex
poles. (d) $y = -2.65$. The poles $\alpha_2, \alpha_{2}^*$ move away from
the real $\alpha$-axis, quickly becoming high-energy resonances.
}\label{patterns}
\end{figure}
Correspondingly, at small negative $\xi^2$ the zero $i(\pi+\alpha_2)$ appears
close to the right from the $A_1$ pole (Fig.7a). The equation \eqref{r1pert}
remains valid at negative $\xi^2$; the fact that the residue at the $A_1$
pole becomes negative is consistent with the non-unitarity of the theory at
$\xi^2 <0$.

As $\xi^2$ continues to decrease further from zero, the parameter
$B_2=\sin\alpha_2$, already negative at $\xi^2 < \xi_{2}^2$, continues to
decrease as well, and at certain value $\xi_{22}^2$ it crosses the value
$-1$. Although below $\xi_{2}^2$ $\ M_2$, and hence $B_2$, can not be
extracted directly from the TFFSA data, it can be estimated using certain
dispersion relation \cite{zyl}. This yields the numerical value of
$\xi_{22}^2$, which we quote in terms of the variable $y$ (Eq.\eqref{yrays})
as well as $\xi^2$,
\begin{eqnarray}
y_{22}=-4.46(4) \qquad\quad (\xi_{22}^2 = -0.0036(2))\,.
\end{eqnarray}
At $\xi_{22}^2$ the poles $i\alpha_2$ and $i(-\pi-\alpha_2)$ collide at
$-i\pi/2$, and at $\xi^2 < \xi_{22}^2$ they move away from the imaginary
$\theta$-axis as a pair of self cross-conjugated complex poles. The patterns
of real poles of $S(\theta)$ slightly above and slightly below $\xi_{22}^2$
are shown in Figs 7b and 7c, respectively. Finally, as $\xi^2$ approaches
$-\xi_{0}^2$ from above, the poles $(i\alpha_2,i\alpha_{2}^*) =
(-i\pi/2+\beta_2,-i\pi/2-\beta_2)$ depart to infinity\footnote{It is possible
that additional resonance poles exist in this domain. In any case, as $\xi^2$
approaches $-\xi_{0}^2$, by scaling arguments all the resonance poles are
expected to move to infinity as $\alpha_p \simeq -\frac{5i}{6}\,
\log(\xi^2+\xi_{0}^2)$.}, and the only poles of $S(\theta)$ left are the s-
and u- channel $A_1$ poles at $2\pi i/3$ and $\pi i/3$ (Fig.7d).
Simultaneously, as $\xi^2 \to \xi_{0}^2 +0$ the discontinuities across the
inelastic branch cuts disappear\footnote{The discontinuities across the
inelastic branch cuts go down as fast as $(\xi^2+\xi_{0}^2)^\frac{14}{3}$, as
we argue in \cite{zyl}.}, and in the limit $S(\theta)$ reduces to
\eqref{syl}.

\section{Resonances}

Although in this work I am not going to discuss behavior of the resonance
poles in any systematic way, here some comments on this issue will be made.

Clearly, the amplitude $S(\theta)$ generally has some number of resonance
poles. We have already observed that the $A_3$ and $A_2$ poles at $i\alpha_3$
and $i\alpha_3$, together with their cross-poles, become self
cross-conjugated complex poles when $\xi^2$ gets below certain values
$\xi_{33}^2$ and $\xi_{22}^2$, respectively. But certainly there may be more
resonance poles. For instance, it was already mentioned that the integrable
theory at $\eta=0$ involves eight stable particles which we refer to as
$A_1$, $A_2$, ..., $A_8$. Only the first three of these particles, $A_1$,
$A_2$ and $A_3$, remain stable when $\eta$ is shifted away from the
integrable point $\eta=0$. The other five particles $A_4, A_5, ..., A_8$
loose stability becoming resonances, and their masses $M_4, M_5, ..., M_8$
acquire imaginary parts, as in \eqref{mpres}.  In fact, for the lowest of
them, $A_4$ and $A_5$ the widths $\Gamma_4$ and $\Gamma_5$ were computed to
the order $\eta^2$ via form-factor perturbation theory in $\eta$
\cite{e8decay}. Below I attempt to follow the fate of some of the associated
complex poles at negative $\eta$, until the poles leave the principal sheet
of the Riemann surface of $S(\theta)$.

Let me start with the following (well known) observation. At $\eta=0$ the
mass $M_3$ of $A_3$ appears to be located very close below the threshold
$2M_1$, i.e. the difference
\begin{eqnarray}\label{epsilon2}
\varepsilon_2 = 2 M_1-M_3 = 4M_1\,\sin^2 \frac{\pi}{60} \approx 0.0109562\ M_1
\end{eqnarray}
is numerically small in the units of $M_1$. This means that at $\eta=0$ the
particle $A_3$ can be interpreted as a weakly coupled two-particle $A_1 A_1$
bound state. It is also well known that in $1+1$ dimensions neutral particles
that form two-particle bound state also tend to form three- four- and
generally $k$-particle bound states, with the binding energies
\begin{eqnarray}\label{epsilonk}
\varepsilon_k \simeq \frac{k^3-k}{3!}\,\varepsilon_2\,,
\end{eqnarray}
where $\varepsilon_2$ is the binding energy of the two-particle bound state.
This is because when the particles are weakly bound, the system is well
approximated by non-relativistic Bose particles with attractive
delta-function interaction, from which \eqref{epsilonk} follows
\cite{deltagas0,deltagas1}. Indeed, typical momenta of the particles
constituting a weakly coupled bound state are small in the units of $M_1$,
making it possible to use non-relativistic theory. Also, the size of the
bound state is much greater then the effective interaction range ($\sim
M_{1}^{-1}$), allowing to approximate the pairwise interactions by
delta-function potentials. One can show that direct multi-particle
interactions become negligible in the limit when the binding energy goes to
zero. Note that even if $\varepsilon_2$ is small, this approximation applies
only to finitely many lowest $k$-particle bound states since at sufficiently
large $k$ the energies \eqref{epsilonk} become comparable with $M_1$, and the
approximation breaks down.

The above argument suggests, in particular, that at $\eta=0$ we have to
observe weakly coupled three- , four- , and perhaps multi - particle bound
states of $A_1$, with the binding energies approximately given by
\eqref{epsilonk}. And indeed, as it turns, $M_5$ is numerically close to
$3M_1$, $M_7$ is close to $4M_1$, and $M_8$ is close to $5M_5$. Using exact
mass ratios \cite{e8}, let us compute the binding energies and compare them
to the approximation \eqref{epsilonk},
\begin{eqnarray}
&&3M_1 - M_5 = \left(3-4\,\cos\frac{\pi}{5}\,
\cos\frac{2\pi}{15}\right)\,M_1 =\,\,0.043704\ M_1\,, \quad
\varepsilon_3 = 0.043824\ M_1\,,\nonumber\\
&&4 M_1 - M_7 = \left(4-8\,\cos^2 \frac{\pi}{5}\,\cos\frac{7\pi}{30}\right)\,M_1 =
0.108843\ M_1\,, \quad
\varepsilon_4 = 0.109562\ M_1\,,\label{epsilons}
\\
&& 5 M_1 - M_8 = \left(5-8\,\cos^2 \frac{\pi}{5}\,
\cos\frac{2\pi}{15}\right)\,M_1 = 0.216613\ M_1\,, \quad
\varepsilon_5 = 0.219124\ M_1\,.\nonumber
\end{eqnarray}
I would like to emphasize that at $\eta=0$, while the mass spectrum of
\cite{e8} is exact, \eqref{epsilonk} is only an approximation -- after all,
$\varepsilon_2$ in \eqref{epsilon2} is only numerically small. Nonetheless,
numerical agreement between the exact mass spectrum and the weak binding
approximation \eqref{epsilonk} is remarkably good - curiously enough, the
Perron-Frobenius vector of the Cartan matrix of $E_8$ "knows" about weakly
interacting particles. Also note that $\varepsilon_5$ in \eqref{epsilons} is
already comparable to $M_1$, so for greater values of $k$ one should not
expect the above arguments to be too reliable.

Consider small nonzero $\eta$. Now the integrability of the theory
\eqref{ift} is broken, and the particles $A_4$, $A_5$, ..., $A_8$ become
unstable against decays into the lighter particles. Correspondingly, the
masses $M_4, M_5, ..., M_8$ immediately become complex, with small imaginary
parts $\Gamma_n \sim \eta^2$.  Although all five particles $A_4$, $A_5$, ...,
$A_8$ turn into resonances, below we argue that the widths of the resonances
$A_5$, $A_7$, $A_8$ remain small at all $\eta$ between $0$ and $\eta_3$, and
go back to zero as $\eta$ approaches $\eta_3$. When $\eta$ decreases below
$\eta_3$ these three particles likely cease to exist even as the resonance
states (the corresponding poles leave the principal sheet of the Riemann
surface of $S(\theta)$). It is plausible that in a similar manner, the widths
of the resonances $A_4$ and $A_6$ go to zero when $\eta$ approaches $\eta_2$
from above, and below $\eta_2$ these resonances disappear.

As was already discussed in Sect.3, as $\eta$ becomes negative, the parameter
$\alpha_3$, already as small as $\pi/15$ at $\eta=0$, moves even closer to
zero. As long as it remains positive, the particle $A_3$ remains stable, but
its mass $M_3 = 2M_1\,\cos\frac{\alpha_3}{2}$ approaches $2M_1$, and its
description as the weakly coupled bound state of two $A_1$ particles with the
binding energy
\begin{eqnarray}
\varepsilon_2 \simeq 4M_1\,\sin^2\frac{\alpha_3}{2}
\end{eqnarray}
become yet more accurate. Therefore there is even better reason to expect the
presence of weakly coupled $k$-particle bound states with $k$ greater then 2
when $\eta$ gets closer to $\eta_3$. Moreover, Eq.\eqref{epsilonk} gives the
better approximations of the binding energies the closer $\alpha_3$ to zero
is. Of course, at nonzero $\eta$ exact integrability is violated, and the
bound states with $k \geq 3$ are unstable against decays into lighter
particles. However, it is possible to argue that the decay rates of the
$k$-particle bound states are suppressed at least as $\varepsilon_{2}^{k-1}$,
therefore these resonances become very narrow ($\Gamma \sim
(\eta-\eta_3)^{2(k-1)}$) as $\eta$ approaches $\eta_3$ from above. When
$\eta$ crosses $\eta_3$ the resonance poles merge with inelastic thresholds,
and then likely move on to further sheets of the Riemann surface of
$S(\theta)$. Below $\eta_3$ the particles $A_5$, $A_7$, $A_8$ disappear even
as the resonance states, at least in the definition given in Sect. 2. On the
basis of the above argument I predict that the width of the resonances $A_5$,
$A_7$, $A_8$ remain very small in the domain $\eta_3 < \eta < 0$, and their
masses in this domain are very well approximated by the equation
$\left(k-(k^3-k)\alpha_{3}^2/24\right)\,M_1$, $k=3,4,5$. It would be
interesting to test this prediction against numerical data. Existing TFFSA
data is consistent with the prediction, but substantial improvements (mostly
in proper handling of the finite size effects) are needed in order to make
meaningful comparison.

There are several questions related to the above argument. If the binding
energy is sufficiently small, one expects to have $k$-particle bound states
with any $k$. But at $\eta=0$ we do not see bound states with $k > 5$.
Perhaps at this point the binding energy is not sufficiently small. However
as $\eta$ decreases, and the binding energy becomes smaller, the higher bound
states have to appear. Where they could come from? At this time I do not have
good answer to this question. I could only speculate that their emergence can
interfere with the formations of bound states of heavier resonances. Note in
this connection that if the weak binding approximation \eqref{epsilonk} with
$k=6$ is applied at $\eta=0$, the resulting binding energy $\varepsilon_6 =
0.383467\ M_1$ would be numerically close to the difference $6 M_1 - M_4 -M_6
=0.3769\ M_1$.

Another question concerns the fate of the resonances $A_4$ and $A_6$, which
were missed in the above scenario. Their $\eta=0$ masses \cite{e8}
\begin{eqnarray}
&&M_4=4 M_1\,\cos\frac{\pi}{5}\, \cos\frac{7\pi}{30} =2.404867\ M_1 \,,\nonumber\\
&&M_6 = 4 M_1\,\cos\frac{\pi}{5}\,\cos\frac{\pi}{30} =3.218340\ M_1
\end{eqnarray}
do not seem to suggest any weakly coupled bound state structure. At small
$\eta \neq 0$, of course they too become resonances. Then, apart from the
opening of narrow decay channel(s), there are no reasons to expect dramatic
structural changes in these particles at small nonzero $\eta$. And indeed,
TFFSA data show the presence of the $A_4$ and (somewhat less convincingly)
$A_6$ resonances at $\eta$ well below $\eta_3$. However the ratios ${\bar
M}_4/M_1$ and ${\bar M}_6/M_1$ (${\bar M}_n$ are the real parts of the masses
$M_n$, Eq.\eqref{mpres}) increase as $\eta$ decreases from $0$. Recall that
as $\eta$ approaches $\eta_2$ (Eq.\eqref{eta2}) from above, the mass $M_2$
gets close to the threshold $2 M_1$ -- now $A_2$ becomes weakly coupled $A_1
A_1$ bound state. Therefore, at $\eta$ slightly above $\eta_2$ one expects to
have a number of narrow resonances with the (complex) masses close to $k
M_1$, $k=3,4,...$, and with the "binding energies" well approximated  by
\eqref{epsilonk}, where now $\varepsilon_2=2M_1-M_2$. It is tempting to
speculate that as $\eta$ decreases to the values slightly above $\eta_2$, the
resonances $A_4$ and $A_6$ assume the role of the weakly coupled three- and
four- particle bound states of $A_1$. This would suggest that at these values
of $\eta$ the imaginary parts $\Gamma_4,\ \Gamma_6$ become small, while the
real parts ${\bar M}_4$ and ${\bar M}_6$ appear closely below $3M_1$ and
$4M_1$, respectively. Again, the existing TFFSA data seems to be consistent
with this scenario, although more detailed analysis is desirable.

\bigskip
\section*{Acknowledgments}

I would like to thank Sergei Lukyanov, Alexei Tsvelik, and especially Barry
McCoy for many invaluable discussions and interest. I gratefully acknowledge
hospitality of the Simons Center for Geometry and Physics where parts of this
work were done.

\bigskip

\noindent The research is supported in part by DOE grant DE-FG02-96ER40959.

\end{document}